%% file: ALPINIST.tex
\documentclass[11pt,a4paper]{article}
\pdfoutput=1
\usepackage{jheppub}
\usepackage{rotating}
\usepackage{array}
\usepackage{amsmath}
\usepackage[normalem]{ulem}
\usepackage{slashed}
\usepackage{booktabs}
\usepackage[pdftex,table]{xcolor}
\usepackage{units}
\usepackage{enumerate}
\usepackage{upgreek}

\newcounter{rowcntr}[table]
\renewcommand{\therowcntr}{\roman{rowcntr}}
\newcolumntype{N}{>{(\refstepcounter{rowcntr}\therowcntr)}r}

\usepackage{multirow}

\arxivnumber{2201.05170, TTK-22-04, CP3-22-02}

\title{ALPINIST: Axion-Like Particles In Numerous Interactions Simulated and Tabulated}

\author[1]{Jan Jerhot,}
\author[2]{Babette D{\"o}brich,}
\author[3]{Fatih Ertas,}
\author[3]{Felix Kahlhoefer}
\author[4]{and Tommaso Spadaro}

\affiliation[1]{Centre for Cosmology, Particle Physics and Phenomenology (CP3), Universit{\'e} catholique de Louvain, Chemin du Cyclotron 2,
B-1348 Louvain-la-Neuve, Belgium}
\affiliation[2]{CERN, Esplanade des Particules 1, 1211 Geneva 23, Switzerland}
\affiliation[3]{Institute for Theoretical Particle Physics and Cosmology (TTK), RWTH Aachen University, D-52056 Aachen, Germany}
\affiliation[4]{Laboratori Nazionali di Frascati INFN, Via E. Fermi 40,
00044 Frascati, Italy}

\emailAdd{jan.jerhot@cern.ch}
\emailAdd{babette.dobrich@cern.ch}
\emailAdd{ertas@physik.rwth-aachen.de}
\emailAdd{kahlhoefer@physik.rwth-aachen.de}
\emailAdd{tommaso.spadaro@cern.ch}

\abstract{Proton beam dump experiments are among the most promising strategies to search for light and feebly interacting states such as axion-like particles (ALPs). The interpretation of these experiments is however complicated by the wide range of ALP models and the multitude of different production and decay channels that can induce observable signals. Here we propose a new approach to this problem by separating the calculation of constraints and projected sensitivities into model-independent and model-dependent parts. The former rely on extensive Monte Carlo simulations of ALP production and decays, as well as estimates of the detection efficiencies based on simplified detector geometries. Once these simulations have been performed and tabulated, the latter parts only require simple analytical rescalings that can be performed using the public code \texttt{ALPINIST} released together with this work. We illustrate this approach by considering several ALP models with couplings to Standard Model gauge bosons. For the case of ALPs coupled to gluons we show that the sensitivity of proton beam dump experiments can be extended significantly by considering hadronic ALP decays into three-body final states.}


\begin{document}
\maketitle
\flushbottom

\section{Introduction}
\label{sec:introduction}

In recent years the search for physics beyond the Standard Model (SM) has begun to expand its focus from particles too heavy to be produced in laboratory experiments to light particles with extremely weak interactions~\cite{Agrawal:2021dbo}. As grows the range of models that predict such light degrees of freedom, so does the number of proposed and ongoing experiments that can probe the corresponding parameter spaces. A central challenge arising from this development is to identify those experiments that promise the greatest improvement in sensitivity to a wide variety of models. A popular way to address this challenge is to define a number of benchmark scenarios, for which exclusion bounds and sensitivity projections are to be calculated~\cite{Beacham:2019nyx,Aielli:2019ivi,Kelly:2020dda}. As the field matures, however, more flexible and model-independent methods are needed that allow for the reinterpretation of experimental bounds and sensitivities for a wide range of different scenarios.

This need is particularly great in the context of axion-like particles (ALPs), which are hypothetical elementary pseudoscalar particles arising as the pseudo-Nambu-Goldstone modes of an approximate global symmetry that is broken both spontaneously at some large energy scale $\Lambda$ and explicitly by the non-zero ALP mass $m_a \ll \Lambda$. ALPs emerge
naturally in many theories extending the SM symmetry group, which may feature new accidental global symmetries that are not experimentally observed and therefore assumed to be broken spontaneously~\cite{Andreas:2010ms}. A plethora of unobserved global symmetries also appears in theories with extra dimensions when these are compactified to four dimensions~\cite{Cicoli:2012sz}. From a phenomenological point of view, the relation $m_a \ll \Lambda$ furthermore implies that ALPs can be viable dark matter (DM) candidates or mediate the interactions between SM and DM particles~\cite{Nomura:2008ru,Batell:2009di,Freytsis:2009ct,Dolan:2014ska}.

Since the details of the spontaneous symmetry breaking and the corresponding new physics at the energy scale $\Lambda$ are typically unknown, ALPs can manifest themselves in a wide range of different ways at low energies. To approach this problem in a general way, one can express the interactions between ALPs and SM particles in terms of an effective field theory valid at energies $E \ll \Lambda$~\cite{Georgi:1986df,Chala:2020wvs,Bauer:2020jbp}. This approach makes it possible to experimentally search for ALPs in a model-independent way, i.e. without referring to a specific symmetry breaking mechanism. While traditionally much energy has been focused on the effective ALP-photon coupling~\cite{Dolan:2017osp}, recent studies have broadened the picture and considered also effective couplings to the other SM gauge bosons~\cite{Izaguirre:2016dfi,Brivio:2017ije,Alonso-Alvarez:2018irt,Aloni:2018vki,Ertas:2020xcc} as well as to SM Higgs bosons~\cite{Bauer:2017ris} and SM fermions~\cite{Cornella:2019uxs,Bauer:2021mvw}.

Ideally, one would like to study all of these effective couplings simultaneously, which requires the production of on-shell ALPs and the observation of their subsequent decays back into SM particles. For $m_a$ in the MeV--GeV range, the most suitable laboratory for this purpose are particle accelerators. While colliders like the LHC have the advantage of larger ALP production cross sections~\cite{Mimasu:2014nea}, fixed-target experiments benefit from higher statistics and lower backgrounds and are therefore more sensitive to tiny couplings and small ALP masses~\cite{Dobrich:2018jyi}. To improve our understanding of ALPs and ultimately achieve their discovery, it is therefore essential to understand the signatures of ALPs in fixed-target experiments for general effective interactions.

In this work we propose a new approach for evaluating experimental sensitivities, which is based on the observation that the calculation of the predicted number of ALP events in a given experiment can be split into a number of {\it separate steps, each of which can be performed in a largely model-independent way}. Specifically, we
\begin{itemize}
    \item write the differential ALP production cross section  as the sum of several different production modes, each of which depends in a well-defined way on the effective ALP couplings;
    \item make use of the fact that the detection probability depends trivially on the various ALP branching ratios, with only the total ALP lifetime having a more complicated effect.
\end{itemize}
This makes it possible to perform all relevant calculations in advance and tabulate them in terms of a small number of model parameters. For a given combination of effective ALP couplings, these tables can then be evaluated and combined in a straight-forward way.

In addition to this conceptual innovation, our work also expands on the range of effects included in our calculations in order to extend existing approaches (using for example the automated simulations of beam-dump experiments in Maddump~\cite{Buonocore:2018xjk}) and to allow for a wide range of applications. On the ALP production side, we consider Primakoff production (both from on-shell~\cite{Dobrich:2019dxc} and off-shell~\cite{Dobrich:2015jyk} photons), ALP-meson mixing~\cite{Bauer:2017ris} and rare decays (both from $B$ mesons~\cite{Freytsis:2009ct} and $D$ mesons~\cite{Carmona:2021seb}). On the ALP decay side, we include for the first time detailed experimental sensitivities for leptonic final states and three-body hadronic final states like $\pi^+ \pi^- \pi^0$. For the latter we have implemented an automatic re-weighting of the Dalitz plot density based on detailed analytic calculations of hadronic decay modes~\cite{Aloni:2018vki,Cheng:2021kjg}. All these calculations are applied to a wide range of past, present and proposed experiments, for which we have implemented a modelling of detector geometries and efficiencies. To illustrate the flexibility of our approach, we apply it to several different ALP models featuring various effective couplings to SM gauge bosons.

Together with this work we release a public code named \texttt{ALPINIST}~\cite{jan_jerhot_2022_5844011}, which provides the relevant tables for ALP production and detection together with helpful routines for model-dependent calculations and a convenient user interface.

The remainder of this work is structured as follows.  In section~\ref{sec:framework} we introduce the general framework used to obtain model-independent constraints on ALP models. The various ALP production and decay modes are discussed in sections~\ref{sec:production} and~\ref{sec:decay}, respectively. A detailed discussion of the experiments considered in this work is given in section–\ref{sec:exp}. To illustrate our approach, we present experimental sensitivities for a range of ALP scenarios in section~\ref{sec:results}. The code \texttt{ALPINIST} is described in more detail in appendix~\ref{sec:ALPINIST}. The calculation of the various effective couplings and the resulting decay widths are discussed in the appendices~\ref{sec:calculations} and~\ref{sec:decay_calculation}.

\section{General framework}
\label{sec:framework}

In this section we outline our general approach for simplifying the calculation of exclusion limits and sensitivity projections for ALPs in a model-independent way. For this purpose we consider a pseudoscalar particle $a$ with mass $m_a$ that couples to the various particles of the SM via effective operators. Limiting ourselves to the case of dimension-5 operators, the strength of each interaction is parametrised by an effective coupling that can be expressed in terms of the new-physics scale $\Lambda$ and the Wilson coefficients $C$ of the various effective operators (see appendix~\ref{sec:calculations} for details). For example, the coupling of the ALP to a pair of photons is given by
\begin{equation}
\mathcal{L}_{a\gamma\gamma} = e^2 \frac{C_{\gamma\gamma}}{\Lambda} a F^{\mu\nu}\tilde{F}_{\mu\nu} \; ,
\end{equation}
where $F^{\mu\nu}$ denotes the electromagnetic field strength tensor and $\tilde{F}^{\mu\nu}$ its dual.
We denote the set of all couplings that define a specific parameter point of a given ALP model by $\mathbf{C}$. More details will be provided in section~\ref{sec:results} and appendix~\ref{sec:calculations}.

The focus of the present work is on proton beam fixed-target experiments, which aim to produce ALPs using a beam of highly energetic protons and detect the subsequent decays of ALPs into SM particles in a downstream detector. At such experiments ALPs can be produced via various mechanisms (production channels), for example through scattering of beam protons or secondary particles on the target material or through the decay of secondary particles.  At first sight, the number $N_\text{det}$ of expected ALP events in the detector depends in a complicated non-linear way on the couplings $\mathbf{C}$, which determine the production cross section, the kinematic distributions, the lifetime $\tau_a = 1 / \Gamma_a$ and the branching ratios into the various final states. Upon closer inspection, it however becomes clear that this quantity can be factorized as follows:
\begin{equation}
 N_\text{det} = \int \mathrm{d}\theta_a \mathrm{d}E_a \frac{\mathrm{d}^2N\left(m_a, \mathbf{C}\right)}{\mathrm{d}\theta_a \mathrm{d} E_a} p_\text{det}\left(m_a, \Gamma_a, \theta_a, E_a, \mathbf{C}\right) \; .
\end{equation}
Here $\mathrm{d}^2N/\mathrm{d}\theta_a\mathrm{d}E_a$ denotes the spectrum of ALPs produced in the target as a function of the angle (relative to the incident beam) and the ALP energy in the laboratory frame, and $p_\text{det}$ denotes the probability that an ALP with given angle and energy will lead to an observable signal in the detector, which depends on the position of the ALP decay as well as on the kinematic distribution of the decay products.

ALP production in the target can proceed via a number of different processes, which will be discussed in detail in section~\ref{sec:production}. The ALP spectrum is therefore simply a sum over all production channels:
\begin{equation}
\frac{\mathrm{d}^2N\left(m_a, \mathbf{C}\right)}{\mathrm{d}\theta_a \mathrm{d} E_a} = \sum_i \frac{\mathrm{d}^2N_i\left(m_a, \mathbf{C}\right)}{\mathrm{d}\theta_a \mathrm{d} E_a} \; .
\end{equation} 
Now the crucial observation is that for each individual production channel, the spectrum depends in a trivial way on the ALP coupling structure, in the sense that changing the couplings only affects the normalisation but not the shape of the spectrum. In other words, it is possible to calculate the spectrum for a given reference coupling $\mathbf{C}_\text{ref}$ and then perform an appropriate rescaling:
\begin{equation}
\frac{\mathrm{d}^2N_i\left(m_a, \mathbf{C}\right)}{\mathrm{d}\theta_a \mathrm{d} E_a} = f_i(\mathbf{C}, \mathbf{C}_\text{ref}) \frac{\mathrm{d}^2N_i\left(m_a, \mathbf{C}_\text{ref}\right)}{\mathrm{d}\theta_a \mathrm{d} E_a}\; .
\end{equation}
For example, if ALPs are produced via their coupling to photons, the production cross section is proportional to $C_{\gamma\gamma}^2$ and hence
\begin{equation}
f_{\gamma\gamma}(\mathbf{C}, \mathbf{C}_\text{ref}) = \frac{C_{\gamma\gamma}^2}{C_{\gamma\gamma,\text{ref}}^2} \; .
    \label{eq:ggg}
\end{equation}
This simple formula holds even if the ALP effective photon coupling receives contributions from several different effective operators, see appendix~\ref{sec:calculations} for details and Ref.~\cite{Bauer:2017ris} for a comprehensive overview of such contributions.

We emphasize that the  ALP mass, on the contrary,  affects the ALP spectrum in a non-trivial way. It is therefore necessary to calculate the spectra $\mathrm{d}^2N_i/\mathrm{d}\theta_a \mathrm{d} E_a$ separately for each ALP mass under consideration. For perturbative production processes such as photon fusion the ALP spectra can be derived from first principles. In many cases, however, production proceeds via non-perturbative processes and the spectra can be only obtained via simulations. These simulations however only need to be performed once (for the chosen values of $\mathbf{C}_\text{ref}$) and can then be applied to any ALP model. 

Similarly, the detection probability can be split into different final states $f$:
\begin{equation}
 p_\text{det}\left(m_a, \theta_a, E_a, \mathbf{C}\right) = \sum_f \text{BR}_{a \to f}(m_a, \mathbf{C}) p_{\text{det},f}(m_a, \Gamma_a, \theta_a, E_a) \; ,
\end{equation}
where $\text{BR}_{a \to f} = \Gamma_{a \to f} / \Gamma_a$ denotes the branching ratio into the final state $f$. While these branching ratios may depend in a non-trivial way on the couplings $\mathbf{C}$, the detection probability $p_{\text{det},f}(m_a, \Gamma_a, \theta_a, E_a)$ for a given final state depends on the couplings only through the ALP lifetime, which determines the probability of the ALP to decay in a given part of the detector. Again, this makes it possible to calculate the detection probabilities $p_{\text{det},f}$ in advance and then apply an appropriate rescaling through the branching ratios $\text{BR}_{a \to f}$.

Putting everything together, we can therefore write
\begin{align}
 N_\text{det} & = \sum_{i,f} f_i(\mathbf{C}, \mathbf{C}_\text{ref}) \text{BR}_{a \to f}(m_a, \mathbf{C}) \times \int \mathrm{d}\theta_a \mathrm{d}E_a \frac{\mathrm{d}^2N_i\left(m_a, \mathbf{C}_\text{ref}\right)}{\mathrm{d}\theta_a \mathrm{d} E_a}  p_{\text{det},f}(m_a, \Gamma_a, \theta_a, E_a) \\
 & = \sum_{i,f} M_{if}(m_a, \mathbf{C}) \times E_{\text{det},if}(m_a, \Gamma_a) \; ,
 \label{eq:Ntot}
\end{align}
where $M_{if}$ denotes the part that depends on the ALP model (i.e.\ on the couplings $\mathbf{C}$), while $E_{\text{det},if}$ encapsulates all the experimental details. Note that in the final line we have omitted the dependence on $\mathbf{C}_\text{ref}$, which will be kept implicit in the following.

As will become clear in the subsequent sections, the calculation of $E_{\text{det},if}$ for a given experiment can be quite challenging, typically requiring Monte Carlo (MC) simulations of both the production and the decay. However, these simulations only have to be performed once (for given $m_a$ and $\Gamma_a$ and a choice of $\mathbf{C}_\text{ref}$) and the (tabulated) functions $E_{\text{det},if}$ can easily be made publicly available. This makes it possible to perform fast analyses of any ALP model for which the model-dependent functions $M_{if}$ have been calculated. Moreover, simulation upgrades including new production mechanisms or decay channels or more realistic experiment modelling can be quickly included in the framework, allowing straightforward estimates of the related phenomenological consequences. 

\section{ALP production}
\label{sec:production}

In this work we consider three types of ALP production channels, which are described in detail in the remainder of this section: production via the effective coupling of ALPs to photons (section~\ref{sec:production:photon}), ALP mixing with neutral pseudoscalar mesons (section~\ref{sec:production:mixing}) and ALPs produced in heavy meson decays (section~\ref{sec:production:decays}). Each of these production channels is efficient for production of ALPs in a different mass range and results in a different $E_a$-$\theta_a$ distribution, thus favouring a different experimental setup. For example, ALPs produced via photon-fusion tend to be forward-emitted and soft; those produced in heavy meson decays can be emitted at large angles and with higher energies.

As discussed in the previous section, the differential ALP yield for each of the production channels can be split into a model-dependent and a model-independent part. For production via the ALP-photon coupling, the model-dependent part is given by eq.~(\ref{eq:ggg}). For production via meson mixing, the corresponding expression is
\begin{equation}
    f_{P}(\mathbf{C}, \mathbf{C}_\text{ref}) = \left| \frac{\theta_{aP}}{\theta_{aP,\text{ref}}}\right|^2 \; ,
\end{equation}
where $P = \pi^0, \eta, \eta'$ and $\theta_ {aP}$ denotes the corresponding ALP-meson mixing angle (see below). Finally, for the production via $B$ meson decay we have
\begin{equation}
    f_{BK^{(\ast)}}(\mathbf{C}, \mathbf{C}_\text{ref}) =  \frac{\text{BR}(B \to K^{(\ast)} + a)}{\text{BR}(B \to K^{(\ast)} + a)_\text{ref}} \; ,
\end{equation}
where
\begin{equation}
    \text{BR}(B \to X) = \frac{\Gamma_{B \to X}}{\Gamma_B} \;,
    \label{eq:BRBK}
\end{equation}
denotes the $B$ meson branching ratios, and analogous expressions for $D$ meson decays.

\subsection{Photo-production}
\label{sec:production:photon}
The ALP photo-production, also called Primakoff process, can be realized in two ways at a beam-dump setup: the electro-magnetic (EM) scattering of a beam proton with a target nucleus and the interaction of on-shell photons from decays of secondary neutral pseudoscalar mesons with the nuclear EM field. The former process can be in turn either realized through the scattering of the whole proton on the target nuclei (elastic scattering) or through the scattering of their constituents. The elastic scattering, which dominates at beam energies of current beam dump facilities, has been studied in details in Ref.~\cite{Dobrich:2015jyk}. In Ref.~\cite{Dobrich:2019dxc} it has been shown that ALP photo-production via on-shell photons from meson decays  constitutes an even a more important source of ALPs. In this work, we include both elastic scattering and Primakoff production from on-shell photons.

In both cases the ALP differential yield can be expressed as
\begin{equation}
\label{eq:primakoff}
\frac{\mathrm{d}^2N}{\mathrm{d} \theta_a \mathrm{d} E_a} = \frac{C_{\gamma\gamma}^2}{\Lambda^2} \frac{8 e^4}{\sigma_N} \frac{\sin \theta_a}{\pi E_{\text{beam}}} \int \text{d} p_t^2 \text{d} \phi_a \frac{\text{d}\sigma_{\gamma N}}{\text{d} \cos \theta_a} \gamma(E_a,p_t^2)\;,
\end{equation}
where the differential cross section for the photon-nucleus interaction is taken from Eq.~(3.16) of Ref.~\cite{Dobrich:2015jyk} and scales with $Z_{\text{target}}^2$. This cross section has to be normalized to the total cross section for the incident particle. In the case of elastic proton scattering, the total cross section for the incident particle is $\sigma_N \simeq 53\,\mathrm{mb} \times A_{\text{target}}^{0.77}$. For the absorption of an on-shell photon, we use values from Ref.~\cite{nist:xcom}.

The photon distribution $\gamma$ can be calculated using the Weizs\"{a}cker-Williams approximation \cite{BUDNEV1975181} in the case of the proton elastic scattering.\footnote{It has been pointed out in Ref.~\cite{Harland-Lang:2019zur} that this approximation becomes inaccurate for small ALP masses as well as for low beam energies. Given that elastic scattering only gives a sub-dominant contribution compared to ALP production from photons produced in meson decays, we find that this approximation is nevertheless fully sufficient for our purposes.} In order to obtain the $\gamma$ distribution for on-shell photons from $\pi^0,\eta,\eta^\prime$ di-photon decays, we have to simulate the production of these secondary mesons. In this work the meson yields from $p$-$p$ collisions at experimentally relevant energy values\footnote{For 400, 120 and 70 GeV proton beams the respective $p$-$p$ c.m. energies are approximately 27.4, 15 and 11.5 GeV.} were simulated using \texttt{PYTHIA 8.2}~\cite{PYTHIA82} with \texttt{SoftQCD:all} flag and a pomeron flux parametrization \texttt{SigmaDiffractive:PomFlux(5)}, a setup which has been validated in Ref.~\cite{Dobrich:2019dxc}. The corresponding photon distribution is then obtained from interpolation of the dataset resulting from the flat phase-space decay of these mesons into photon pairs. The decay products are boosted to the laboratory frame and the events are weighted with the $\pi^0,\eta,\eta^\prime$ di-photon branching fractions.

\subsection{ALP production from meson mixing}
\label{sec:production:mixing}

If the ALP field is coupled to the SM quark and/or gluon fields, it naturally enters the kinetic and mass matrices of the effective Lagrangian of pseudoscalar mesons. Upon diagonalization of these matrices in order to obtain the mass eigenstates, the mixing angle between ALPs and mesons $P = \pi^0,\eta,\eta^\prime$ emerges in the following form:
\begin{equation}
\label{eq:thetaMix}
\theta_{aP} = \frac{f_{\pi}}{f_a} \frac{K_{aP} m_a^2+ m_{aP}^2}{m_a^2-m_P^2}\;,
\end{equation}
where the ALP decay constant $f_a$ depends on the underlying ALP couplings. We emphasize that this expression is valid for values $m_a$ such that $\left\vert\theta_{aP}\right\vert^2 \ll 1$. For the explicit form of $f_a$, the kinetic mixing terms  $K_{aP}$ and the mass mixing terms $m_{aP}$ we refer to appendix~\ref{sec:calculations} and Ref.~\cite{Aloni:2018vki}.

In the presence of such a mixing it is possible to produce an ALP instead of the meson $P$ via SM processes in a small fraction of cases proportional to $|\theta_{aP}|^2$. The corresponding differential yield can be written as
\begin{equation}
\frac{\mathrm{d}^2 N}{\mathrm{d} \theta_a \mathrm{d} E_a} = \vert \mathcal{F}_{\text{VMD}}(m_a) \vert^2 \sum_P \frac{\hat{\alpha}^2_s(m_a)}{\hat{\alpha}^2_s(m_P)} \frac{\mathrm{d}^2N_P}{\mathrm{d} \theta_P \mathrm{d} E_P } \bigg|_{E_P \to E_a \atop \theta_P \to \theta_a }  \left\vert\theta_{aP} \right\vert^2,
\end{equation} 
where $\mathrm{d}^2N_P/\mathrm{d}\theta_P \mathrm{d}E_P$ denotes the differential yield of the meson $P$. To consider ALP masses up to $3\,\mathrm{GeV}$ and capture the suppression of ALP production for $m_a \gg m_P$, we follow Ref.~\cite{Aloni:2018vki} and include the vector meson dominance (VMD) form factor
\begin{equation}
\label{eq:Fvmd}
    \mathcal{F}_{\text{VMD}} (m) = \left\{ \begin{array}{ll} 
        1, & \text{for } m \leq 1.4\,\mathrm{GeV}\\
        \sum\limits_{i = 0}^3 a_i m^i, & \text{for } 1.4\,\mathrm{GeV} < m \leq 2\,\mathrm{GeV}\\
        (1.4\,\mathrm{GeV}/m)^{4}, & \text{for } 2\,\mathrm{GeV} < m \leq 3\,\mathrm{GeV}\\
        0, & \text{for } m>3\,\mathrm{GeV},
        \end{array}\right.
\end{equation}
and the running strong coupling $\hat{\alpha}_s (m)$ in the form
\begin{equation}
\label{eq:alphas}
    \hat{\alpha}_s (m) = \left\{ \begin{array}{ll} 
        1, & \text{for } m \leq 1\ \text{GeV}\\
        \sum\limits_{i = 0}^3 b_i m^i, & \text{for } 1\,\mathrm{GeV} < m \leq 1.5\,\mathrm{GeV}\\
        4 \pi / 7 \log (m / 0.34\,\mathrm{GeV})^2, & \text{for } m > 1.5\,\mathrm{GeV} ,
        \end{array}\right.
\end{equation}
where parameters $a_i$ and $b_i$ are determined by requiring continuity of the resulting function and its first derivative. Note that compared to Ref.~\cite{Kelly:2020dda} we assume a stronger suppression of the ALP yield for large ALP masses to ensure that our estimates are conservative.

For obtaining the distribution of $\mathrm{d}^2N_P/\mathrm{d}\theta_P \mathrm{d}E_P$, we can conveniently re-use the validated datasets from the simulation described in section~\ref{sec:production:photon}. There is however an important subtlety: Since the pseudoscalar meson masses are generally different from the ALP mass, it is not possible to simply replace a pseudoscalar by an ALP with the same energy and momentum. Even if the replacement is made in such a way that the three-momentum is conserved in a specific frame, it may not be conserved in a different frame. This introduces some degree of arbitrariness in the precise prescription used for the replacement. Here we require that the three-momentum of the ALP be equal to the three-momentum of the pseudoscalar meson in the $p$-$p$ \emph{cms frame}, calculate the corresponding ALP energy and then boost the result into the laboratory frame. For further details we refer to appendix~\ref{ap:mixing_kinematics}.

\begin{figure}[t]
\begin{center}
\includegraphics[scale=0.73]{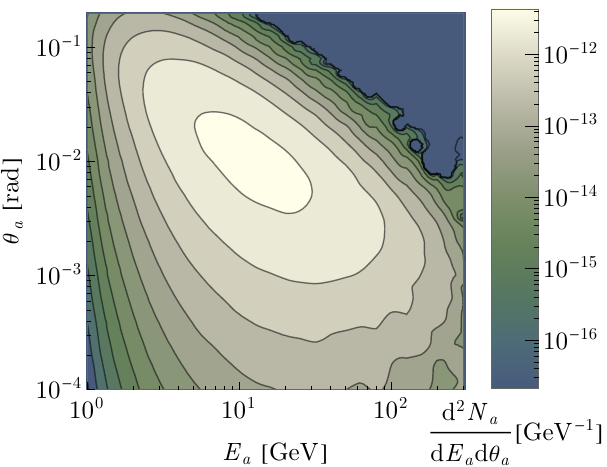}
\includegraphics[scale=0.73]{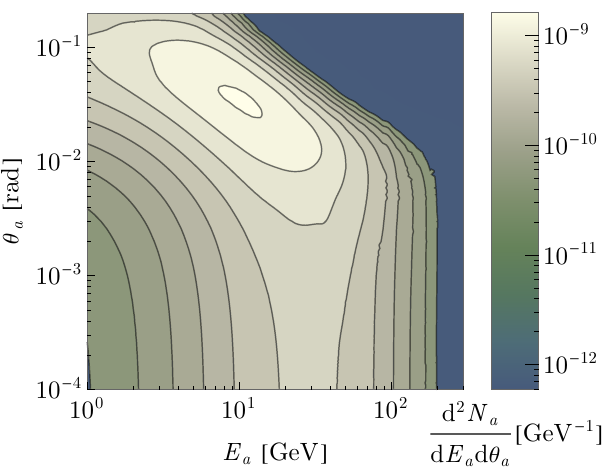}
\caption{\label{fig:prod_meson} Production differential yields for an $m_a = 0.5$ GeV ALP produced at a copper target with a 400 GeV proton beam. For a \textit{photon-from-meson} induced production (left) with a fixed $C_{\gamma\gamma} / \Lambda = 10^{-4}\,\mathrm{GeV}^{-1}$ and for an $a$-$\eta$ mixing (right) assuming a fixed mixing angle $\theta_{a\eta} = 10^{-4}$. 
} 
\end{center}
\end{figure}

In figure~\ref{fig:prod_meson} we show the predicted distribution of ALPs in the $E_a$-$\theta_a$ plane for the photon-from-meson production channel (left panel) and for production via ALP-$\eta$ mixing (right panel). In both cases we find an anti-correlation between the two parameters in the sense that higher ALP energies imply smaller production angles and vice versa. For the meson mixing contribution, the differential yield peaks at larger ALP angles than for Primakoff production, but still gives a sizeable contribution in typical on-axis beam-dump experiments with an angular acceptance of a few mrad.

\subsection{ALP production from rare decays}
\label{sec:production:decays}

Flavour-changing ALP couplings arise not only from interactions of ALPs with quarks~\cite{Dolan:2014ska} and electroweak gauge bosons~\cite{Izaguirre:2016dfi} but also 
in higher-order processes from ALP-gluon interactions~\cite{Chakraborty:2021wda}. In all of these cases ALPs can be produced in flavour-changing neutral current (FCNC) processes and can therefore be probed through rare meson decays in fixed-target experiments  \cite{Ertas:2020xcc,Dobrich:2018jyi}. However, in order for the produced ALPs to be focused in the forward direction it is essential that the parent meson decays before interacting with the target material. $B$ and $D$ mesons can thus be of particular importance since when produced at the beam energies of $\mathcal{O}(100)\,\mathrm{GeV}$ their decay length is significantly smaller than the interaction length of the target material.\footnote{The nuclear collision length for most of the target materials is $\approx 10$ cm \cite{Zyla:2020zbs}. The decay length for a $100\,\mathrm{GeV}$ momentum $B$ meson is $< 1$ cm and $\approx 1\,\mathrm{cm}$ for a $D$ meson, while for $K$ it is $> 500\,\mathrm{cm}$.}

Let us first consider the case of $B$ mesons. In this work we include the ALP production from $B \to K^{(\star)} + a$ decays (both charged and neutral). Since the decay is isotropic in the rest frame of the $B$ meson, the differential ALP yield depends only on the distribution of $B$ mesons and the branching ratios for the decays involving ALPs (see eq.~(\ref{eq:BRBK})).
The $B$ meson distributions are again obtained using simulations of $p$-$p$ collisions with \texttt{PYTHIA 8.2}. We follow the setup of Ref.~\cite{Dobrich:2018jyi}, i.e.\ we allow only the bottom quark production hard-QCD processes in order to have sufficient statistics and then reweight the final distribution according to the bottom production cross section at the given beam energies. The value reported by \texttt{PYTHIA} is $\sigma_{pp} = 39.85 \times 10^{9}\,\mathrm{pb}$ and $\sigma_{bb} = 1.866 \times 10^{3}\,\mathrm{pb}$ for $400\,\mathrm{GeV}$ beam, which has already been validated in Ref.~\cite{Dobrich:2018jyi} using the values reported in Ref.~\cite{Lourenco:2006vw}. For $120$ (and $70\,\mathrm{GeV}$) the values are $\sigma_{pp} = 38.54 \times 10^9\,\mathrm{pb}$ and $\sigma_{bb} = 3.8\,\mathrm{pb}$ ($\sigma_{pp} = 38.38 \times 10^9\,\mathrm{pb}$ and $\sigma_{bb} = 1.15 \times 10^{-7}\,\mathrm{pb}$ respectively), for which no measurement has been found in the existing literature for validation. In any case, for experiments operating with such small beam energies the ALP production via $B$ meson decays is found to be negligible (see section~\ref{sec:results}). A separate simulation involving the two-body decay kinematics of $B \to K^{(\star)} + a$ is performed to obtain the resulting ALP distribution as a function of $E_a$ and $\theta_a$.

In an analogous way, we simulate the production of $D$ mesons with 
\begin{equation}
    \sigma_{cc} = \begin{cases} 
        3.601 \times 10^{6}\,\mathrm{pb}, & \text{for } E_{\text{beam}} = 400\,\mathrm{GeV}\\
        0.518 \times 10^{6}\,\mathrm{pb}, & \text{for } E_{\text{beam}} = 120\,\mathrm{GeV}\\
        1.551 \times 10^{5}\,\mathrm{pb}, & \text{for } E_{\text{beam}} = 70\,\mathrm{GeV}
        \end{cases}
\end{equation}
and a subsequent decay $D \to \pi + a$. Despite a larger production cross section of $D$ mesons, for the ALP models considered in this work the branching ratio $\text{BR}_{D \to \pi + a}$ is many orders of magnitude smaller than $\text{BR}_{B \to K^{(\star)} + a}$ as discussed in appendix \ref{sec:calculations}. Nevertheless, we enclose the $D$ mesons data sets to allow also studies of specific ALP models enhancing the up-type quark transition (see e.g.~Ref.~\cite{Carmona:2021seb}).

\begin{figure}[t]
\begin{center}
\includegraphics[scale=0.73]{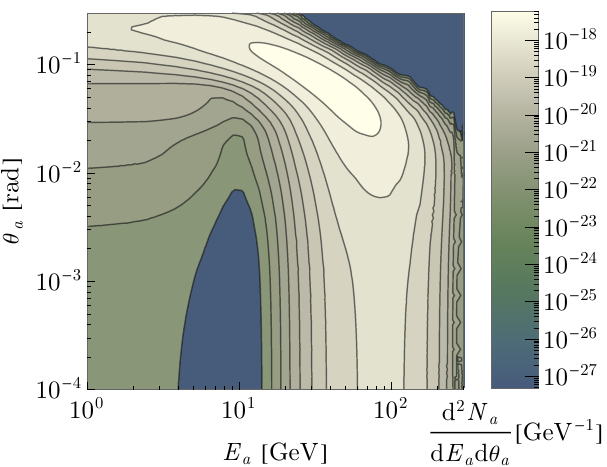}
\includegraphics[scale=0.73]{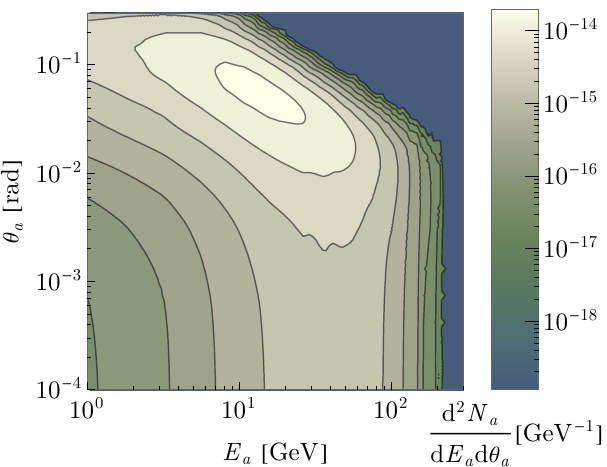}
\caption{\label{fig:prod_BD} Production differential yield of an $m_a = 0.5\,\mathrm{GeV}$ ALP produced in decays $B^{\pm,0} \to K^{\pm,0} + a$ (left) and $D^{\pm,0} \to \pi^{\pm,0} + a$  (right) of $B$ and $D$ produced at a copper target with $400\,\mathrm{GeV}$ proton beam assuming a referential value $\text{BR}_{B \to K + a} = \text{BR}_{D \to \pi + a} = 10^{-10}$.}
\end{center}
\end{figure}

The ALP distribution resulting from $B \to K + a$ and $D \to \pi + a$ decays are plotted in figure \ref{fig:prod_BD}. Compared to the distributions shown in figure~\ref{fig:prod_meson} we observe that the ALP distribution from rare meson decays peaks at even larger angles and energies.  In particular ALPs produced from $B$ mesons have very high energies, making this production mode particularly promising for off-axis experiments searching for relatively heavy ALPs with short lifetimes.

\section{ALP decay and detection of final states}
\label{sec:decay}

\begin{figure}[t]
\begin{center}
\includegraphics[scale=1]{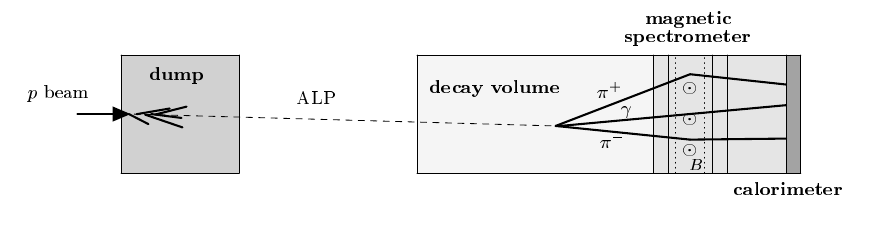}
\caption{\label{fig:scheme} Schematic view of the experimental setup for the case that the ALP decays via $a \to \pi^+ \pi^- \gamma$.}
\end{center}
\end{figure}

To determine the probability that an ALP induces an observable signal, we have modelled the various fixed-target experiments under consideration in a  simplified\footnote{For example we do not include veto detectors in the simulation and we do not account for efficiencies of individual detectors and possible scattering or absorption of final state particles in the material of the experiment before they are detected.} and generic MC simulation of ALP propagation and decays. The layout consists of a decay volume, a spectrometer, analyzing magnets and a forward calorimeter (see schematic drawing in figure~\ref{fig:scheme}). Our set-up is described in detail in section~\ref{sec:dec:fw}.

The outcome of the simulation is the function $E_{\text{det},if}(m_a, \Gamma_a)$ defined in eq.~(\ref{eq:Ntot}). We provide tabulated data sets of $E_{\text{det},if}$ based on a simulation with $N=10^6$ trials on a grid of $\Gamma_a$ and $m_a$ values for each of the production and decay channels $i$,$f$, and for each of the experiments described in section \ref{sec:exp}. We note that the function $E_{\text{det},if}$ includes the efficiency $\epsilon_{\text{det},f}$ of the experiment for the detection of a given final state $f$. For past experiments we use the efficiencies as determined by the experiment (see section \ref{sec:exp}), while for future experiments we simply set $\epsilon_{\text{det},f} = 1$. However, it is straight-forward in our approach to redo our analysis for different values of  $\epsilon_{\text{det},f}$ making our framework a powerful tool to assess the impact of the experimental design on sensitivity projections.

\subsection{General principles of the ALP Monte Carlo simulation}
\label{sec:dec:fw}

The probability $p_{\text{det},f}$ that an ALP with given $m_a$, $\Gamma_a$, $\theta_a$ and $E_a$ decaying to channel $f$ will be detected at the experiment is a potentially complicated function of the distance $l$ that the particle travels before decaying and the momenta of the final-state particles, which can be characterised by a set of kinematical variables $a_1,.. a_n$. The conditions imposed on the decay products by experimental cuts  can be summarized by a function
$\Theta_{\text{det},f}(m_a, \theta_a, E_a,  l, a_1,.. a_n)$, which is equal to 1 if all experimental conditions are met and 0 otherwise. Since it is typically impossible to perform the integration over all of these variables, 
it is useful to define $p_{\text{det},f}$ instead as a sum over randomly generated decay processes
\begin{equation}
p_{\text{det},f}(m_a, \Gamma_a, \theta_a, E_a) = \sum\limits_{k=1}^N \epsilon_{\text{det},f} \frac{ \Theta_{\text{det},f}(m_a, \Gamma_a, \theta_a, E_a; \left\lbrace  l, a_1,.. a_n\right\rbrace_{\text{ran}}^k ) }{N},
\end{equation}
where $\left\lbrace l, a_1,.. a_n \right\rbrace_{\text{ran}}^k$ denotes the k-\textit{th} simulated event of the ALP decay. 

To carry out these simulations we first calculate the ALP decay length in the laboratory frame $l_a = \gamma_a \beta_a / \Gamma_a$ with $\gamma_a = E_a / m_a$ and $\beta_a = p_a / m_a$ and use the distribution $p(l) = l_a^{-1} \exp(-l/l_a)$ to choose a random value of $l$. 
For the decay itself we generate the four-momenta of the second generation particles, as well as extra sets for the third generation particles if the former include unstable particles (the specific distributions used to simulate the ALP decays will be discussed in section~\ref{sec:dec:modes}). Given the magnetic fields of the experiment, these four-momenta enable us to calculate the trajectories of the final-state particles and check the experimental conditions in order to evaluate $\Theta_{\text{det},f}$. 

In practice, in order to obtain the distribution $E_{\text{det},if}(m_a, \Gamma_a)$ for given values of $m_a$ and $\Gamma_a$ we first read the distribution $\Delta^2N_i/\Delta\theta_a \Delta E_a$ for the production mode $i$ from the previous section into a $\theta_a$-$E_a$ histogram  using \texttt{ROOT}~\cite{ROOT}. At each iteration we sample a random bin $\left\lbrace \theta_a, E_a \right\rbrace_{\text{ran}}^k$ according to the histogram and simulate the ALP decay for these values. By summing up a sufficiently large sample $N$, we then obtain a reliable estimate of $E_{\text{det},if}(m_a, \Gamma_a)$ in the form
\begin{equation}
E_{\text{det},if}(m_a, \Gamma_a) = \sum\limits_{k=1}^N \epsilon_{\text{det},f} \frac{ \Theta_{\text{det},f}(m_a; \left\lbrace \theta_a, E_a , l, a_1,.. a_n\right\rbrace_{\text{ran}}^k)}{N} \; .
\end{equation}
Note that the right-hand side depends implicitly on $\Gamma_a$ through the distribution used to obtain $l$. Although the ALP detection probability depends sensitively on the ALP energy $E_a$ and angle $\theta_a$, the left-hand side of the equation above is independent of these quantities and only depends on the chosen production mechanism (represented by the index $i$). In other words, $E_{\text{det},if}$ can be thought of as the expectation value of the detection probability for the distribution of ALP energies and angles predicted by the chosen production mechanism.

\subsection{Decay channels}
\label{sec:dec:modes}

Since beam dump experiments can probe ALPs of masses up to $\mathcal{O}(1)\,\mathrm{GeV}$, there are a number of kinematically allowed decay channels, listed in table~\ref{tab:decay_channels}. We restrict ourselves to two- or three-body decays, therefore accounting for the dominant contribution to the total decay width for the ALP masses and coupling scenarios under consideration (see section~\ref{sec:results}). 

\begin{table}[t]
\begin{center}
\begin{tabular}{ l|c|c }
\hline
 & \textit{neutral} & \textit{charged} \\ \hline
 \multirow{2}{*}{2-body} & $2\gamma$ & $e^+e^-$ \\  
 &  & $\mu^+\mu^-$ \\ \hline 
 \multirow{3}{*}{3-body}  & $3\pi^0$ & $\pi^+\pi^-\gamma$ \\
 & $2\pi^0\eta$ & $\pi^+\pi^-\pi^0$\\ 
 & & $\pi^+\pi^-\eta$ \\   \hline
\end{tabular}
\caption{Allowed decay channels for ALPs with masses up to about $1\,\mathrm{GeV}$.\label{tab:decay_channels}}
\end{center}
\end{table}

For the purposes of the MC simulation, we consider charged particles to be stable and $\pi^0$ and $\eta$ to decay instantaneously into a $\gamma$-$\gamma$ pair with the appropriate branching fraction. 
While two body decays of ALPs are treated with a flat phase space, for three-body decays a more careful treatment is necessary.

In a three-body decay, the invariant masses of first and second and second and third particles, $m_{12}$ and $m_{23}$, can be used to define a Dalitz plot.\footnote{In all of the decays that we consider, there are at least two particles with the same mass. These are taken to calculate $m_{12}$.}
If the transition amplitude square $\vert \mathcal{M} \vert^2$ of the process depends on the momenta of outgoing particles, the resulting Dalitz plot density is generally non-uniform. Such a momentum dependence arises in a wide range of models, for example if the decay proceeds via a resonance, as in the case of $a \to \pi\pi\gamma$, which receives a contribution from virtual $\rho$ meson exchange.

In this work, for the simulation of hadronic decays, we employ the framework derived in Ref.~\cite{Aloni:2018vki} using chiral perturbation theory. The results from Ref.~\cite{Aloni:2018vki} are derived under the assumption that ALPs are coupled only to gluons, but they remain valid also for ALPs with flavour-universal couplings to light quarks. In such a case, the decay width $\Gamma_{a \to \text{hadrons}}$ changes but the ratio $\vert \mathcal{M} \vert^2_{a \to \text{hadrons}} / \Gamma_{a \to \text{hadrons}}$, which determines the distribution of the final-state hadrons, remains invariant. However, for models with non-universal couplings to light quarks, as considered for example in Ref.~\cite{Cheng:2021kjg}, there may be significant differences in both the individual branching ratios and the corresponding distributions of the decay products.

\begin{figure}[t]
\begin{center}
\includegraphics[scale=0.57]{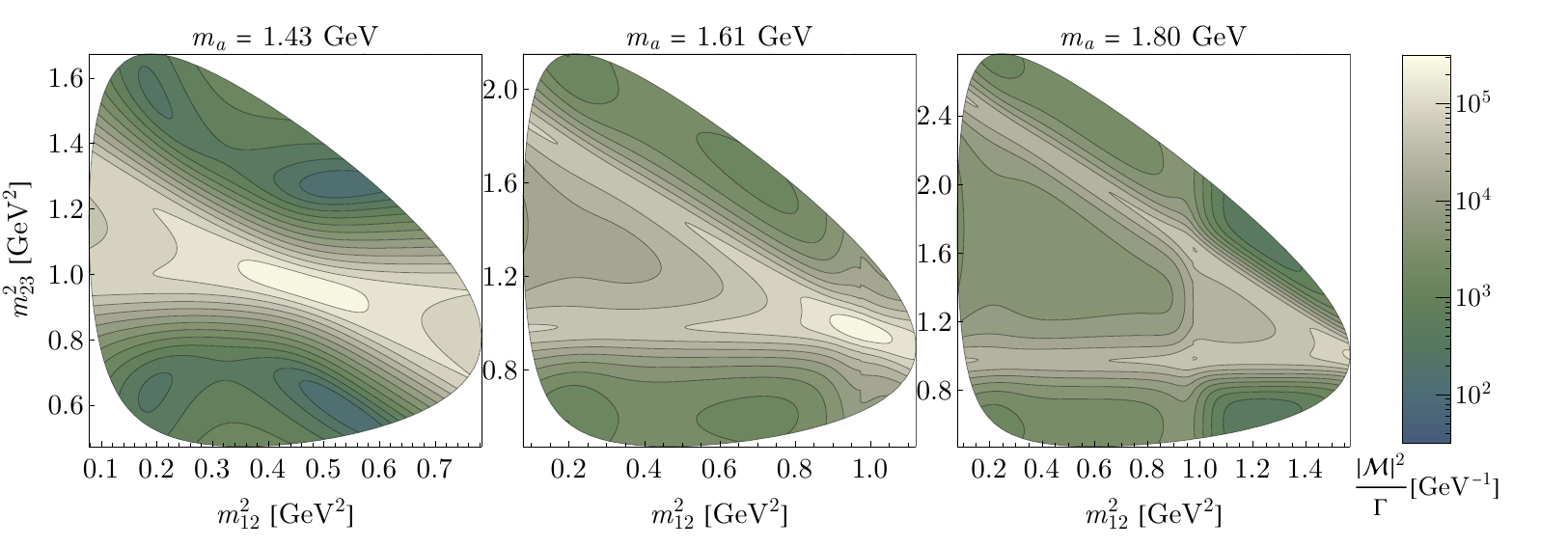}
\includegraphics[scale=0.57]{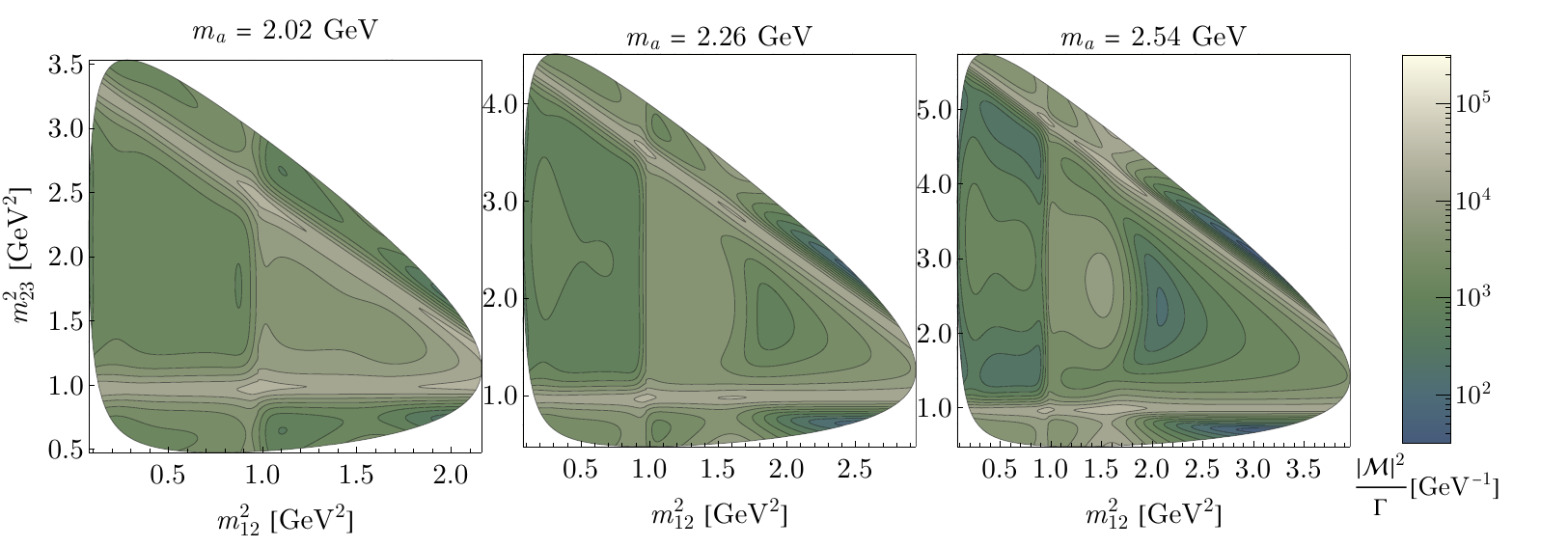}
\caption{\label{fig:dalitz_reweight} Dalitz plot density for six values of ALP masses used for the $a \to \pi^+\pi^-\eta$ decay.}
\end{center}
\end{figure}

For a given ALP mass $m_a$, we use $\vert \mathcal{M} \vert^2_{a \to \text{hadrons}} / \Gamma_{a \to \text{hadrons}}$ from Ref.~\cite{Aloni:2018vki} to weight a flatly distributed Dalitz-plot density. The procedure is repeated for each $m_a$ bin.
As an example, the density for the decay $a \to \pi^+\pi^-\eta$ is shown in  figure~\ref{fig:dalitz_reweight}. 

\subsection{Propagation of decay products}
The photons, either emitted as daughter particles of the ALP or as tertiary decay products, are propagated in straight line up to the upstream face of the forward calorimeter. When extrapolating any charged particle, we account for the bending induced by the spectrometer analyzing magnets, if present: we treat the field as that of an ideal solenoid and do not model effects due to detailed field maps, stray fields, etc. 

Photons and $e^{\pm}$ are assumed to release all their energy in the calorimeter. No resolution effects are simulated and any experimental condition is directly applied on the photon or $e^{\pm}$ energy. A minimum separation between photon or $e^\pm$ showers is required for ``cluster counting": a pair of energy releases is merged into a single ``cluster" if the relative distance of the calorimeter impact points is below an experiment-dependent radius. In this case, the cluster energy is the energy sum of the two merged particles. For cluster counting, the particle impact point is required to be within the calorimeter sensitive region and an experiment-dependent minimum energy is required. Muons are extrapolated forward up to the front face of a muon detector, if present. For muon detection, the muon impact point at the front face of the muon detector is required to be in the sensitive region. The presence of a passive shield in front of a muon detector is only accounted for by applying a loose minimum-energy threshold.

It should be emphasized that we consider ALP decays into photons not only in the decay volume but also in the spectrometer area, effectively enlarging the decay volume up to the calorimeter location. 

\section{Experimental framework}
\label{sec:exp}

In this section we briefly discuss how we have modelled the experimental setups considered in this work and the related acceptance conditions. Table \ref{tab:exp} provides  an overview of the basic parameters used for each experiment. A beam of protons with energy $E_{\text{beam}}$ is made to interact with an absorber based on the element listed in the column ``Target''. The total number of protons on target (POT) is denoted by $N_\text{POT}$. The experiment should be sensitive to ALPs produced in the target and reaching a decay volume (DV) located $z_{\text{DV}}$ meters downstream. The longitudinal axis of the DV has an angle $\theta_{\text{off}}$ with respect to the beam axis, and $\Omega_\text{cov}$ denotes the solid angle covered by the calorimeter. Neutral (charged) ALP-daughter particles can be detected if the ALP decays within $l_{\text{n(c),DV}}$ meters from the DV entrance. Details are given in the following subsections. 

\begin{table}[tbp]
\resizebox{\columnwidth}{!}{%
\begin{tabular}{ cccccccccc } 
 \hline
 Experiment & Status & $E_{\text{beam}}$ & $N_{\text{POT}}$ & Target & $l_{\text{n,DV}}$ & $l_{\text{c,DV}}$ & $z_{\text{DV}}$ & $\theta_{\text{off}}$ & $\Omega_\text{cov}$ \\ 
   &  & [GeV] & [$10^{18}$] &  & [m] & [m] & [m] & [mrad] & [$\upmu$sr] \\ \hline
 CHARM & completed & $400$ & $2.4$ & Cu & $35$ & $35$ & $480$ & $10$ & $34$ \\ 
 NuCal & completed & $70$ & $1.7$ & Fe & $23$ & $23$ & $64$ & $0$ & $700$ \\  
 NA62 & running & $400$ & $10$ & Cu & $139$ & $81$ & $82$ & $0$ & $84$ \\  
 DUNE ND\footnotemark & proposed & $120$ & $1100$ & C & $10$ & $10$ & $575$ & $0$ & $36$ \\  
 DarkQuest & proposed & $120$ & $1.44$ & Fe & $13.5$ & $1$ & $5$ & $0$ & $12000$ \\ 
 SHADOWS & proposed & $400$ & $10$ & Cu & $26$ & $23$ & $10$ & $75$ & $9200$ \\  
 SHiP & proposed & $400$ & $200$ & Mo & $50$ & $28$ & $45$ & $0$ & $4500$ \\ \hline
\end{tabular}
}
\caption{\label{tab:exp} Overview of the basic parameters for the experimental setups considered in this work. See text for details and definitions.}
\end{table}
\footnotetext{Since the operational time is not yet decided, we evaluate the data sets for one year of data-taking at DUNE. Our data sets can be easily re-scaled and the resulting contours in \ref{sec:results} correspond to ten years of operation. In Ref.~\cite{Kelly:2020dda}, a ten year operation is considered with a larger number of POT integrated per year than that considered here. We assume the DUNE near detector to be operated on-axis. Changing the detector vertical position in our simulation is straightforward and can be performed on request.}

\subsection{Results from past experiments}

To explore future opportunities,
first of all we need to recast the results of the past CHARM and NuCal experiments in terms of their sensitivity for ALPs. 

\subsubsection{CHARM}
For CHARM (using a copper target), an off-axis search for two photons in the final state was presented in Ref.~\cite{CHARM:1985anb}, with an average efficiency of $0.51$. 
The search required at least one photon detected at the calorimeter. The sensitive area of the calorimeter has transverse dimensions of $3 \, \mathrm{m} \times 3\,\mathrm{m}$. 
In the analysis quoted, the shower energy was required to be between $5$ and $50\,\mathrm{GeV}$ and this condition has been implemented in our model.
The same reference includes results
for the detection of final states with two daughter muons.\footnote{This is in contrast to the simulation performed in Ref.~\cite{Dobrich:2018jyi}, where it was mistakenly assumed that CHARM would be sensitive also to final states with just {\it one} muon.} The detection efficiency is reported to be $0.85$.
 In our simulation, the muons are required to have at least $1\,\mathrm{GeV}$ at the final detector plane in order to be seen as minimum ionizing particles throughout the detector material.
 
Regarding hadronic final states, we are only
aware of on-axis searches, which go hand in hand with a larger background rate. 
Ref.~\cite{CHARM:1983vkv} reports around $80$ muon-less events  based on a statistics of $7\times 10^{17}\,\mathrm{POT}$~\cite{CHARM:1980zcj}, after requiring a shower-like energy release of more than $20\,\mathrm{GeV}$. According to the quoted analysis, $4\pm21$ events can be attributed to the  interaction of other neutrinos or  neutrino-like  particles.
Due to the limitation in statistics and due to the uncertainty induced by the subtraction of the background, the strongest exclusion by CHARM on hadronic decay modes is again given by a re-interpretation of the search for decays to final states with emitted photons (Ref.~\cite{CHARM:1985anb}): final states such as $\pi^+\pi^-\gamma$, $\pi^+\pi^-\pi^0$, and $\pi^+\pi^-\eta$ are considered and at least one and at most two photons are required to be detected in the calorimeter, while all of the charged hadrons are required to escape the volume.

\subsubsection{NuCal}
The sensitivity of NuCal to final states with photons was evaluated in Ref.~\cite{Blumlein:1990ay}, requiring the detection of one electromagnetic shower. While NuCal operated with a $70\,\mathrm{GeV}$ proton beam, it benefited from a relatively small distance between target and detector ($64\,\mathrm{m}$) and from a comparably large decay volume length ($23\,\mathrm{m}$). 
After rejecting hadronic-like showers, NuCal observed $1$ event compared to a background expectation of $0.3$ events. ALP signals corresponding to a prediction of more than $3.6$ events are excluded with a confidence level of $90\%$.
For the detection of a two-track final state, we follow the procedure from Ref.~\cite{Blumlein:2013cua} and require both tracks to reach a circle of $2.6\,\mathrm{m}$ diameter at the end of the decay volume with a combined energy of at least $10\,\mathrm{GeV}$.
As in Ref.~\cite{Blumlein:2013cua}, we assume an overall detection efficiency of $0.7$ for the photonic final state, and $0.8$ for the di-muon final state.

By contrast,  we are not aware of corresponding results for hadronic final states in NuCal. Thus, we proceed similarly as for CHARM: we reinterpret the search for final states with photons as appropriate (i.e.\ if the final state contains at least one $\gamma$, $\pi^0$, or $\eta$) and require that the charged hadrons escape detection.

\subsection{Acceptance of current experiments}

The only operational proton beam experiment that we will consider in the following is NA62. Some first data taken in beam dump also exist from the KOTO experiment~\cite{KOTO}. The modelling of KOTO might be easily added to our code. Doing so would however require us to re-run and validate the PYTHIA simulation for a $30\,\mathrm{GeV}$ beam energy, which is non-trivial because of limited available data.

\subsubsection{NA62}

The primary goal of NA62 is the precise measurement of the rare decay of $K^+ \rightarrow \pi^+ \nu \bar{\nu}$, but the experiment is also sensitive to a variety of exotics scenarios.
To search for neutral final states, NA62 has to be run in beam-dump mode, otherwise the residual background can limit the sensitivity.
For a dump-mode run, the standard T10 target of NA62 must be removed, and the movable TAX collimators located $22\,\mathrm{m}$ downstream of T10 must be closed and used to dump the proton beam. This operation has been validated during the 2016--2018 data-taking and about $10^{16}\,\mathrm{POT}$ have been collected in beam-dump mode. Triggers sensitive to decay channels with neutral and charged daughter particles have been deployed. During the current 2021--2023 run $\sim 10^{18}\,\mathrm{POT}$ will be collected in beam dump mode. Around $10^{17}\,\mathrm{POT}$ have been collected in 2021~\cite{PBCNA62Massri}. Here we discuss the implications of collecting data in beam dump mode in the data-taking period between the so-called long-shutdown 3 (LS3) and LS4 to obtain $\sim 10^{19}\,\mathrm{POT}$.\footnote{Note that there are several projects in consideration for NA62 in the post-LS4 period which would allow collection of even larger statistics~\cite{PBCNA62}. These are not considered in this work.} We assume no background limitation, in line with the present knowledge based on 2021 data~\cite{PBCNA62Massri}.

We model NA62 in a toy MC as follows: the beam-defining collimator for `regular' data-taking is used to dump the beam and its distance to the start of the fiducial volume is $82\,\mathrm{m}$. A decay region in vacuum contains a spectrometer with a first station $81\,\mathrm{m}$ downstream of the fiducal volume entrance. The spectrometer hosts a $0.7\,\mathrm{T}$ solenoidal magnet $95\,\mathrm{m}$ downstream of the entrance of the decay volume. The Liquid Krypton Calorimeter (LKr) is located about $139\,\mathrm{m}$ downstream of the entrance of the decay volume and it is modelled as an octagon, see e.g.\ Ref.~\cite{NA62:2017rwk}.

For the neutral final states, we require the following acceptance conditions: Both photons from the ALP decay must be detected at a minimum mutual distance of $10$~cm at the LKr plane. In addition, the energy of each photon must be above $1\,\mathrm{GeV}$ and each photon impact point at the LKr must be at least $15\,\mathrm{cm}$ away from the LKr central hole through which the beam-pipe passes. Finally, the energy sum of the two photons must exceed $3\,\mathrm{GeV}$. The NA62 target material (i.e.\ the material of the upstream section of the final collimator) is copper.

For the di-muon decay, we require both muons to be in the acceptance of the NA62's first and last tracking stations. The acceptance loss due to the central hole allowing the passage of the beam pipe is modelled, cf. also Ref.~\cite{Dobrich:2018jyi}.
Each of the muons should have at least $5\,\mathrm{GeV}$ in order to be tracked efficiently.

Finally, for hadronic final states that include photons, 
not only the $\pi^+$ and $\pi^-$ must be tracked, but also
the additional requirement is made that all photons reach the calorimeter and deposit at least $3\,\mathrm{GeV}$ of combined energy.

\subsection{Acceptance of proposed experiments}

To evaluate the sensitivity of future experiments, we base our analysis on the available proposals. Some of the endeavours we considered are more defined, while others might be subject even to significant change in the future. Moreover, some of the proposed experiments may face non-negligible backgrounds, which we do not attempt to estimate in the present work.

\subsubsection{SHiP}
Since the envisaged geometry for SHiP has changed since the publication of the SHiP proposal~\cite{SHiP:2015vad}, we follow the layout
of Ref.~\cite{Ahdida:2654870} for our estimates.
The prospects for detection of photons from ALP decays in the SHiP calorimeter is modelled as follows. The fiducial region is taken to be $45\,\mathrm{m}$ downstream of the production point of the ALPs and contains a spectrometer with the first spectrometer station located $50.76\,\mathrm{m}$ after the entrance of the decay volume and a $0.15\,\mathrm{T}$ magnet. The calorimeter is situated after the forth spectrometer station\footnote{The exact distance between the calorimeter and the last spectrometer station is not stated in the used literature, so we use an approximate value of $50\,\mathrm{cm}$ inferred from the schematics.}, which is $60.16\,\mathrm{m}$ downstream of the entrance of the decay volume.

We ask both photons to be in an acceptance area of $5\,\mathrm{m} \times 10\,\mathrm{m}$. The energy of each photon must be above $1\,\mathrm{GeV}$. The sum of the photon energies should be above $3\,\mathrm{GeV}$. The photon impact points at the calorimeter should be at least $10\,\mathrm{cm}$ apart. The target material is molybdenum and the POT are $10^{20}$. Note that the proposed SHiP calorimeter has the potential of reconstructing the photon direction, which allows for an ALP mass reconstruction.

To model the detection of charged particles in SHiP we mimic the steps provided in Ref.~\cite{SHiP:2020vbd}.
The decay vertex must lay in the decay volume and the two tracks must hit all spectrometer chambers in their sensitive volume, with a minimum distance of $5\,\mathrm{cm}$ away from the decay vessel walls. Each particle must have a minimum energy of $5\,\mathrm{GeV}$.
Finally, for hadronic final states that include also photons,
 all photons should reach the calorimeter and deposit at least $3\,\mathrm{GeV}$ of combined energy.

\subsubsection{DarkQuest}

To make projections for DarkQuest, we consider the proposed extension of the SeaQuest experiment~\cite{SeaQuest:2017kjt}, particularly the proposed phase-I parameters. Details of the DarkQuest setup can be found in Ref.~\cite{Batell:2020vqn}.
The target material is iron.
Following Ref.~\cite{Berlin:2018pwi},  we assume for the di-photon final state the need to detect 10 signal events to claim a signal beyond the background fluctuations.

It is assumed that the experiment is upgraded with a calorimeter placed between tracking stations 3 and 4 (at $\sim 18.5\,\mathrm{m}$ downstream of the target). The fiducial volume entrance is $5$ meters downstream of the target\footnote{While this work was undergoing finalization, an independent study of the DarkQuest sensitivity to ALPs has appeared \cite{Blinov:2021say}, using $7\text{-}8\,\mathrm{m}$ as fiducial volume (following \cite{Berlin:2018pwi}, whereas our fiducial volume definition follows \cite{Batell:2020vqn}). Apart from this major difference in the acceptance simulation, there
are slight differences regarding the overall required energy and the
photon separation.}. The first spectrometer station is located about $6\,\mathrm{m}$ from the target and a geometric acceptance of $2\,\mathrm{m}\times2\,\mathrm{m}$ is assumed in the transverse direction. We assume using the ``phase-I statistics'' of $1.44 \times 10^{18}\,\mathrm{POT}$. 

As with the other setups, we require a minimum energy of $1\,\mathrm{GeV}$ for each detected photon and a total energy of at least $3\,\mathrm{GeV}$ and -- given the photon shower Moliere radius -- a minimum mutual distance of $10\,\mathrm{cm}$ at the calorimeter plane to avoid shower overlap.
For the di-muon final state,
we model tracking in the simulation using the KMAG magnet and impose that the particles be in the acceptance of the first and third tracking stations. In addition we require $1\,\mathrm{GeV}$ of minimum energy for each muon so that it is not stopped in the iron absorber.\footnote{Note that this requirement goes slightly beyond that made in Ref.~\cite{Batell:2020vqn}, where no minimum energy requirement is made. There, instead, only the transverse $p_T$ kick induced by KMAG and the third tracking station are considered.}
Following Ref.~\cite{SeaQuest:2017kjt}, muon track pairs should be detected with a $\sim 60$\% efficiency for di-muon masses above $4.2\,\mathrm{GeV}$.
Here, we do not consider the possibility to also detect  hadronic final states, as the $K_L$ background expected for the phase-I set-up is expected to be a limiting factor~\cite{Batell:2020vqn}.

\subsubsection{DUNE ND}

Motivated by the recent results of Ref.~\cite{Kelly:2020dda} we also include the DUNE near detector~\cite{DUNE:2021tad} (ND) in our study. DUNE ND uses a neutrino beam produced at the Long Baseline Neutrino Facility (LBNF) at Fermilab by a $120\,\mathrm{GeV}$ proton beam impacting on a graphite target with integrated intensity of $1.1 \times 10^{21}\,\mathrm{POT}$ per year~\cite{DUNE:2016hlj}. We note that the nominal intensity can increase with further updates and the target material can be changed to beryllium~\cite{DUNE:2016hlj}. Since DUNE ND is located underground separated by approximately $0.5\,\mathrm{km}$ of earth from the target, it can also effectively serve as an on-axis\footnote{The DUNE ND will also have the opportunity to take data up to $60\,\mathrm{mrad}$ off-axis as so-called DUNE-PRISM. This option is not considered in this work since the amount of time allocated to off-axis operation is unknown. We would like to stress, however, that since in certain scenarios ALPs in beam dumps are dominantly produced off-axis, DUNE-PRISM can have a large potential for this type of hidden sector physics searches even if operated for a short period of time.} proton beam dump experiment.

The essential parts of DUNE ND used in our analysis are the Liquid Argon detector (ND-LAr) and a multipurpose detector based on gaseous argon (ND-GAr). The ND-LAr is a $5\,\mathrm{m}$ long time projection chamber with height of $3\,\mathrm{m}$ and width of $7\,\mathrm{m}$ located $574\,\mathrm{m}$ downstream of the target. The ND-GAr located right downstream of the ND-LAr is composed of a $5\,\mathrm{m}$ long cylindrical high pressure gaseous time projection chamber of $5.2\,\mathrm{m}$ diameter surrounded by an electromagnetic calorimeter, in a $0.5\,\mathrm{T}$ magnetic field and a muon system. In this analysis we treat the ND-LAr and ND-GAr as two independent decay volumes. Following the reconstruction in Ref.~\cite{DUNE:2021tad}, we allow the ALP decays to happen in $6\,\mathrm{m} \times 3\,\mathrm{m} \times 2 \,\mathrm{m}$ volume of ND-LAr, excluding $50\,\mathrm{cm}$ from the sides and upstream and $150\,\mathrm{cm}$ downstream corresponding to roughly $5\,\mathrm{cm} \times 9\,\mathrm{cm}$ Moliere radius to allow a full reconstruction of the showers. For ND-GAr we assume a $4\,\mathrm{m}$ long, $2\,\mathrm{m}$ high and $4.8\,\mathrm{m}$ wide block located $580\,\mathrm{m}$ from the target and exclude the outer layers following the exclusion in the active volume calculation of Ref.~\citep{DUNE:2021tad}. The photons in the $2\gamma$ decays in the ND-GAr must have a minimum $0.8\deg$ ($14\,\mathrm{mrad}$) separation angle and a minimum $20\,\mathrm{MeV}$ energy each. Charged hadrons from hadronic ALP decays must have a minimum kinetic energy of $5\,\mathrm{MeV}$, based on the measured resolution \citep{DUNE:2021tad}. For decays taking place in the volume of ND-LAr we do not put any conditions on the minimum energy until the expected performance of the detector based on the ProtoDUNE \citep{DUNE:2017pqt} results is known.  

For muons in DUNE, following Figure 2.30 of~\cite{DUNE:2021tad}, we assume
muons with energies below $1\,\mathrm{GeV}$ are reconstructed with full efficiency. Above those energies, the angle w.r.t. to the beam axis must be smaller than $40\deg$ ($700\,\mathrm{mrad}$).

\subsubsection{SHADOWS}

SHADOWS is a proposed off-axis experiment in the NA62 experimental cavern, which can take data concurrently with NA62 operated in beam dump mode. It would use a $400\,\mathrm{GeV}$ proton beam dumped in the copper TAXes and can take up to $10^{19}\,\mathrm{POT}$ in the data-taking period between LS3 and LS4. The layout of the detector we have employed is based on the 2021 PBC proposal \cite{Baldini:2021hfw} and can be subject to changes in the near future. The decay volume starts $10\,\mathrm{m}$ downstream of the target and accommodates a $2.5\,\mathrm{m}$ long spectrometer with four tracking stations. The spectrometer is located $33\,\mathrm{m}$ downstream of the target. The distance among the first two stations is $50\,\mathrm{cm}$. A $1.5\,\mathrm{m}$ gap between the second and the third station is foreseen to accommodate a $1\,\mathrm{T}$ dipole magnet. The fourth station is $50\,\mathrm{cm}$ downstream of the third. A calorimeter is located about $50\,\mathrm{cm}$ behind the fourth spectrometer station. The calorimeter sensitive area has a rectangular shape of $2.5\,\mathrm{m} \times 2.5\,\mathrm{m}$ and in our implementation the detector is off axis by $\theta_{\text{off}} = 75\,\mathrm{mrad}$, so that the calorimeter center is shifted by $2.25\,\mathrm{m}$ from the beam axis. The calorimeter is likely to be succeeded by a muon detector\footnote{We place the muon detector plane $1\,\mathrm{m}$ behind the calorimeter plane.} since muons generated in the TAXes present the main background component based on the preliminary studies in Ref.~\cite{Baldini:2021hfw}. In this analysis we optimistically assume negligible background, which may overestimate the sensitivity.
The request on muons is similar to that of NA62 and SHiP: Two spectrometer chambers should be hit as well as a minimum energy of $5\,\mathrm{GeV}$ for each particle.

\section{Results}
\label{sec:results}

Having obtained the model-independent functions $E_{if}$ by combining the differential ALP yields from section~\ref{sec:production} with the various ALP decay models discussed in section~\ref{sec:decay} and the experimental set-ups considered in section~\ref{sec:exp}, we are now in the position to calculate the predicted number of ALPs in a given experiment for any ALP model for which the model-dependent functions $M_{if} = f_i(\mathbf{C}, \mathbf{C}_\text{ref}) \text{BR}_{a \to f}(m_a, \mathbf{C})$ are known. The necessary formalism has been worked out in great detail for a broad class of ALP effective theories that include interactions of ALPs with SM quarks and leptons, as well as gauge and Higgs bosons. In the present work we will illustrate this procedure for the case that the ALP interacts dominantly with SM gauge bosons:
\begin{equation}
\label{eq:Lgen}
\mathcal{L}_{a,\text{int}} =  g'^2 \frac{C_{BB}}{\Lambda} a B^{\mu \nu} \tilde{B}_{\mu \nu} + g^2 \frac{C_{WW}}{\Lambda} a W^{\mu \nu} \tilde{W}_{\mu \nu} + g_s^2 \frac{C_{gg}}{\Lambda} a G^{\mu \nu} \tilde{G}_{\mu \nu}.
\end{equation}
Here $g'$, $g$ and $g_s$ denote the hypercharge, weak and strong gauge couplings respectively, $X^\mu = B^\mu, W^\mu, G^\mu$ denote the corresponding gauge fields (with the group generator index suppressed), $X^{\mu\nu}$ denotes the field strength tensor and $\tilde{X}^{\mu\nu} = \tfrac{1}{2} \epsilon^{\mu\nu\rho\sigma} X_{\rho \sigma}$ denotes its dual. 

Rather than varying all three couplings independently, we will consider four different benchmark scenarios:

\vspace{3mm}
\begin{tabular}{ N l l l }
\label{item:1} & $\boldsymbol{B}$ \textbf{dominance:} & $C_{BB} \neq 0$; & $ C_{WW} = C_{gg} = 0$\\
\label{item:2} & $\boldsymbol{W}$ \textbf{dominance:} & $C_{WW} \neq 0$; & $ C_{BB} = C_{gg} = 0$\\
\label{item:3} & \textbf{Gluon dominance:} & $C_{gg} \neq 0$; & $ C_{BB} = C_{WW} = 0$\\
\label{item:4} & \textbf{Co-dominance:} & \multicolumn{2}{l}{$C_{BB} = C_{WW} = C_{gg} \neq 0$}\\
\end{tabular}
\vspace{3mm}

\noindent The first three scenarios are chosen to highlight the different production channels. In particular, the first scenario is similar to the frequently studied case of \emph{photon dominance} (see e.g.\ Ref.~\cite{Dobrich:2019dxc}), except that it includes additional interactions between ALPs and $Z$ bosons, which are however of no relevance for fixed-target experiments. The second scenario additionally predicts FCNC processes involving ALPs, effectively enhancing the ALP production. The third scenario features ALP-meson mixing and decays of ALPs into hadronic final states. Finally, the fourth scenario, first proposed in Ref.~\cite{Ertas:2020xcc} and explored further in Ref.~\cite{Kelly:2020dda} investigates the potential interplay between the different couplings.

In order to predict ALP signals in beam dump experiments, one first needs to obtain explicit expressions for the functions $f_i(\mathbf{C}, \mathbf{C}_\text{ref})$, which describe how the contributions from the different production channels depend on the fundamental interactions in eq.~(\ref{eq:Lgen}).  These functions are readily available in the literature and reproduced in appendix \ref{sec:calculations}. The second step is to calculate the ALP decay length and branching ratios as a function of the couplings and the ALP mass. In most cases the dominant decay mode will be the one into a pair of photons, with decays into leptons only appearing at the one-loop level and therefore giving a negligible contribution.\footnote{This conclusion might change when considering direct couplings of the ALPs to leptons. We provide all necessary data sets as well as the implementation in the code to consider such a scenario. Note however that we do not currently consider ALP decays via the Bethe-Heitler process~\cite{Blumlein:1991xh}, which may be relevant in the case of ALPs coupled dominantly to leptons.} For scenarios with $C_{gg} \neq 0$ and $m_a > 3 m_\pi$ there will also be a relevant contribution from three-body decays into hadronic final states. Additional details are provided in appendix~\ref{sec:decay_calculation}.

Once the functions $M_{if}$ have been obtained, it is straight-forward to calculate the predicted number of ALP events $N_{\text{det}}$ for each experiment as a function of the fundamental couplings and the ALP mass. We provide these numbers in tabulated form together with this work for convenience. In the remainder of this section we show the exclusion limits (projected sensitivities) for past (future) experiments at $90\%$ confidence level. We first consider the case that only the di-photon final state can be detected and subsequently explore the improvement in reach that can be achieved when also hadronic three-body final states are considered.

\subsection{Considering only di-photon decays}
\label{sec:res:photons}

For both scenarios (\ref{item:1}) and (\ref{item:2}) one finds $\mathrm{BR}(a \to \gamma\gamma) \approx 1$ since the one-loop processes $a \to \ell\ell$ are suppressed by many orders of magnitude. Moreover, since we assume $C_{gg} = 0$ in both cases, the production via meson mixing vanishes ($f_P = 0$), while $f_{\gamma\gamma} \neq 0$ thanks to the contribution from $C_{BB}$ and $C_{WW}$ to the effective ALP-photon coupling (see appendix \ref{sec:calculations}). The only difference between the two scenarios is that for (\ref{item:1}) there are no FCNCs and hence $f_{BK^{(\star)}} = 0$, whereas they give an important contribution for (\ref{item:2}). Likewise, the contribution of $f_{D\pi}$ vanishes for (\ref{item:1}) and gives a marginal contribution for (\ref{item:2}) (see appendix \ref{sec:calculations}).

\begin{figure}[t]
\begin{center}
\includegraphics[width=\textwidth]{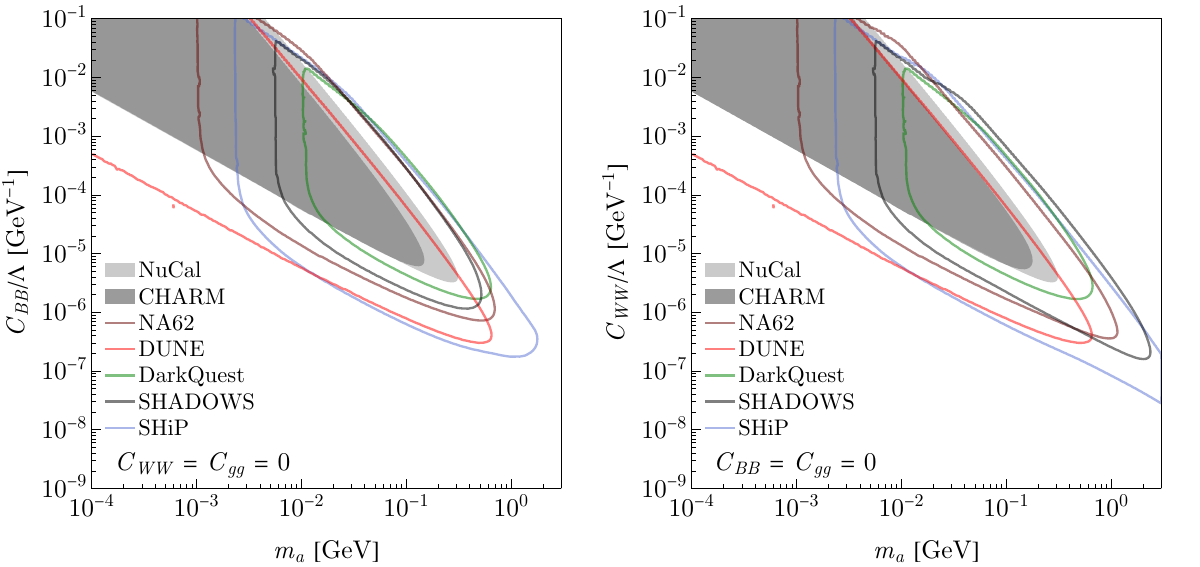}
\caption{\label{fig:excl_BB_WW} 90\% CL exclusions limits for scenarios (\ref{item:1}) (left) and (\ref{item:2}) (right). The exclusion from past experiments which accounts for signal and background studies is indicated by a full contour while the future experiments, assuming a zero background hypothesis, are indicated by a corresponding line. 
}
\end{center}
\vspace{-5mm}
\end{figure}

The existing exclusion limits and projected sensitivities for the two scenarios are compared in figure~\ref{fig:excl_BB_WW}.\footnote{Note that the slight differences in reach compared to Ref.~\cite{Dobrich:2019dxc} can be mainly attributed to the inclusion of $\eta^\prime$ decays and of a wider range of meson momenta.} As expected, all experiments under consideration follow a similar pattern, with the shape of the (projected) exclusions dictated by the ALP decay length (which decreases for large couplings and large ALP masses) and the production cross section (which decreases for small couplings and large ALP masses). Another notable feature in these plots is the low-mass cut-off in sensitivity for most future experiments. This cut-off results from the required separation distance between the two photons in the final state, which decreases for smaller ALP masses and correspondingly larger boost factors.

One finds that in order to extend the reach towards smaller couplings, the decisive quantity is the assumed POT, which is largest for DUNE and SHiP. To make improvements towards larger couplings and masses, the beam energy and detector geometry are decisive. In particular, we observe that the contribution from rare $B$ meson decays in (\ref{item:2}) is especially relevant for experiments with $E_{\text{beam}} = 400\,\mathrm{GeV}$ and clearly favours off-axis experiments like SHADOWS as well as on-axis experiments with large angular coverage like SHiP.

\begin{figure}[t]
\begin{center}
\includegraphics[width=\textwidth]{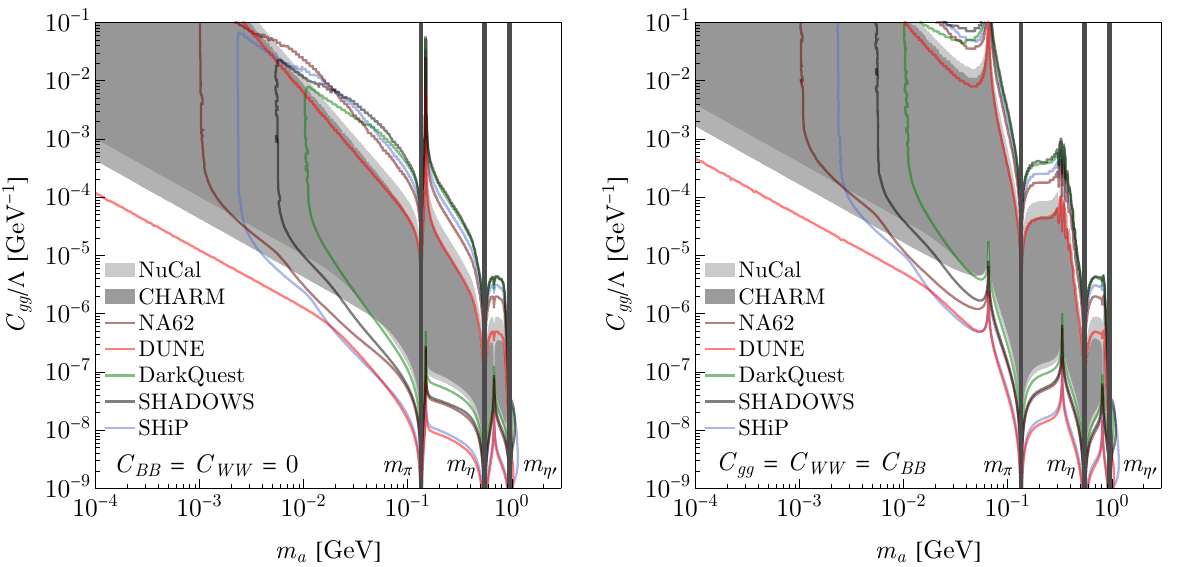}
\caption{\label{fig:excl_GG_COD} 90\% CL exclusions limits for scenarios (\ref{item:3}) (left) and (\ref{item:4}) (right) assuming a di-gamma search only.}
\end{center}
\end{figure}

In scenarios (\ref{item:3}) and (\ref{item:4}) the gluon coupling is non-zero and therefore all production channels considered in this work become relevant.\footnote{Note that we include the two-loop contribution to $B \to K^{(\star)} + a$ from ALP-gluon couplings first calculated in Ref.~\cite{Chakraborty:2021wda}. See appendix~\ref{sec:calculations} for details.} Moreover, one finds $\mathrm{BR}(a \to \gamma\gamma) \ll 1$ for $m_a > 1\,\mathrm{GeV}$ where hadronic decays dominate \cite{Aloni:2018vki}. The effects of ALP-meson mixing in both production and decay is clearly visible in figure~\ref{fig:excl_GG_COD} and imprints a ``pole structure'' in the exclusion plots whenever the ALP mass approaches one of the pseudoscalar meson masses. Note that if the two masses become very close, the condition $\vert \theta_{aP} \vert \ll 1$ no longer holds. These regions are indicated by a dark shading and masked in our analysis. 

In scenario~(\ref{item:4}) there furthermore occurs a partial cancellation in the effective ALP-photon coupling, see eqs.~(\ref{eq:Cgg}) and (\ref{eq:Cggeff}), which has two effects: First of all, for small ALP masses ($m_a \lesssim 100\,\mathrm{MeV}$) it shifts all exclusion contours to somewhat larger couplings, which are necessary to achieve comparable ALP production yields. Second, it leads to additional ``poles''  outside of the masked regions whenever the different contributions to $C_{\gamma\gamma}$ cancel accidentally.

\subsection{Including hadronic decays}
\label{sec:res:hadrons}

\begin{figure}[t]
\begin{center}
\includegraphics[width=\textwidth]{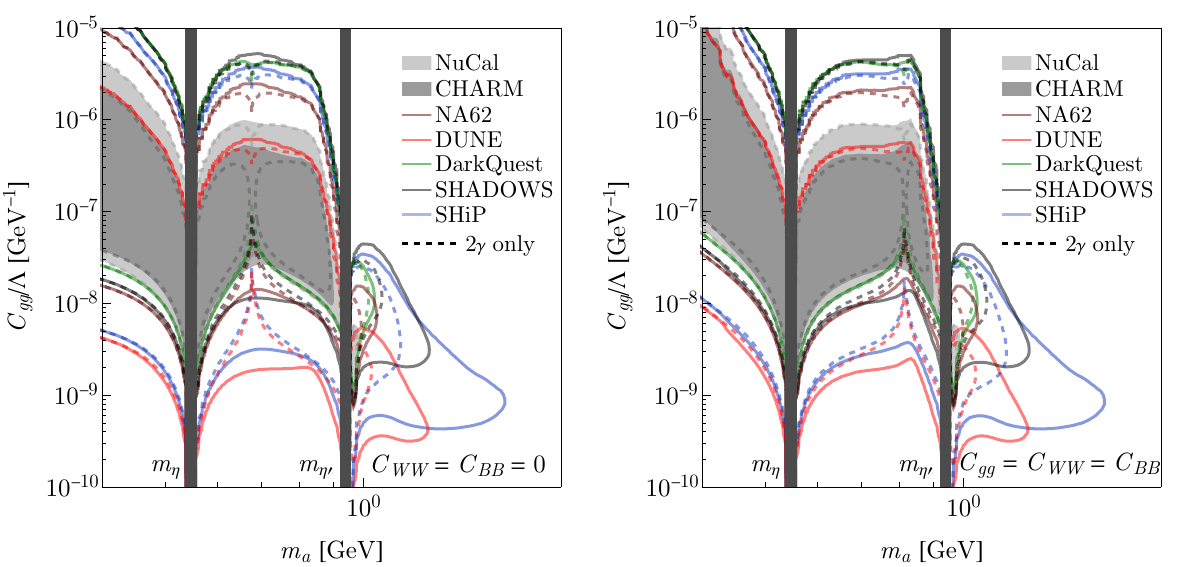}
\caption{\label{fig:excl_tot} 90\% CL exclusions limits for scenarios (\ref{item:3}) (left) and (\ref{item:4}) (right) given by a di-gamma decay only search (dashed line), compared with the exclusion when all mentioned hadronic channels are also taken into account (full line). }
\end{center}
\end{figure}

For scenarios with ALP-gluon couplings the branching ratio into photons is strongly suppressed for heavy ALP masses and whenever there is a cancellation in the effective ALP-photon coupling. The incorporation of hadronic decays is therefore essential to probe these parameter regions and utilize the full potential of beam dump experiments for ALP searches. This can be seen in figure~\ref{fig:excl_tot}, which compares the exclusions and sensitivities for the di-photon channel only and the combination of all channels. We find that the improvement in sensitivity is most significant for SHiP, where the reach in terms of the ALP mass is extended by almost a factor of 2. The reason for this is simply that SHiP is the overall most sensitive experiment and therefore most dependent on the branching ratios of GeV-scale ALPs. Nevertheless, the effect of including hadronic final states is clearly visible also for all other experiments that we consider.

We emphasize that even experiments vetoing hadronic final states can partially cover regions with suppressed decays into photons thanks to hadronic decays with only photons in the final state and with charged hadrons escaping detection (provided that a partial reconstruction of the event is allowed). For experiments sensitive to a range of different final states, on the other hand, our analysis highlights the exciting possibility that beam dump experiments may not only discover ALPs, but also infer their dominant branching ratios, giving us crucial clues regarding the underlying model.

\section{Conclusions}
\label{sec:conclusions}

The search for light feebly interacting particles at experiments with low centre-of-mass energy but high intensity is one of the most exciting frontiers of modern particle physics. A particularly attractive target are axion-like particles (ALPs), for which the smallness of the mass and the weakness of the interactions can simultaneously be motivated by their origin as Pseudo-Goldstone bosons of a spontaneously broken global symmetry. However, the wide range of possible ALP models raises the question how to systematically explore the parameter space and how to present experimental results in a model-independent way.

In this work we have addressed this challenge by developing a novel framework that allows to evaluate constraints and projected sensitivities from proton beam dump experiments for different ALP models without the need to re-run expensive Monte Carlo simulations. This is achieved by splitting the calculation into model-independent parts, which are computationally hard but can be performed and tabulated in advance, and model-dependent parts, which only require simple analytical rescaling functions.

Specifically, we have shown that ALP production can be written as a sum over a number of different production channels, each of which scales in a well-defined way with the effective ALP parameters at low energies. Most of these production channels require Monte Carlo simulations of the hadron shower created by the incident proton in the absorber and the subsequent decays of the resulting mesons. We have performed all necessary simulations using state-of-the-art tools and explicitly validated the resulting cross sections.

Regarding ALP propagation and decay, there are two main difficulties. The first one is to accurately simulate ALP three-body decays, which become important for GeV-scale ALPs with hadronic couplings. The second difficulty is to estimate the detection probability for a given final state in a wide range of experiments. To simulate three-body decays, we have developed a Monte Carlo technique based on the reweighting of Dalitz plots obtained from analytical calculations of hadronic ALP decays in chiral perturbation theory. Detector efficiencies are then estimated by propagating the decay particles through simplified detector geometries, which we have implemented for a range of past, present and future experiments.

To illustrate our approach we present exclusion limits and sensitivity projections for ALPs with couplings dominantly to Standard Model gauge bosons. To illustrate the impact of the detailed coupling structure, we consider four different scenarios, called $B$ dominance, $W$ dominance, gluon dominance and co-dominance. We find crucial differences between these four scenarios regarding the relative importance of different production and decay channels. In particular, for scenarios with gluon couplings, we find that experimental sensitivities can be significantly extended by including hadronic final states from three-body decays in the analysis.

The tabulated simulations and a Python script to perform the necessary rescaling for a user-defined combination of the various ALP interactions with SM particles are publicly available at \url{https://github.com/jjerhot/ALPINIST}. This code allows for the calculation of experimental constraints and sensitivities for a wide range of ALP models. Furthermore, it can be easily extended to include additional experiments, production and decay channels. Further details are provided in appendix~\ref{sec:ALPINIST}.

We emphasize that our implementation is still not fully model-independent. In particular the simulation of three-body decays presently assume the Dalitz plot densities predicted for ALPs with couplings to gluons and/or flavour-independent couplings to SM quarks. Different distributions are expected in more complicated new-physics scenarios, such as the one discussed in Ref.~\cite{Cheng:2021kjg}. We therefore aim to make also the code for Monte Carlo simulations publicly available, such that the user can easily modify decay distributions and experimental geometries. At the same time, we hope that independent Monte Carlo simulations will be developed by the experimental collaborations to produce similar tables to the ones used in our analysis based on more refined detector simulations. This will ultimately allow for a complete reinterpretation of experimental results in the general parameter space of ALP models.

\acknowledgments
We thank Yotam Soreq for useful discussions and Gaia Lanfranchi for comments on our implementation of SHiP and SHADOWS. This  work  is  funded  by  the  Deutsche Forschungsgemeinschaft (DFG) through the Collaborative Research Center TRR 257 ``Particle Physics Phenomenology after the Higgs Discovery'' under Grant  396021762 -- TRR 257 and the Emmy Noether Grant No.\ KA 4662/1-1. This work is also supported through the European Research Council under grant ERC-2018-StG-802836
(AxScale project). JJ is supported through a FRIA grant by the F.R.S.-FNRS (Fonds de  la~Recherche Scientifique - FNRS), Belgium. 

\begin{appendix}

\section{Model-dependent calculations}
\label{sec:calculations}

Although our analysis in section~\ref{sec:results} considers only couplings to $B$ and $W$ bosons and gluons (see eq. (\ref{eq:Lgen})), it is useful to start with a more general effective Lagrangian of ALP-SM interactions:
\begin{align}
\label{eq:Lall}
\begin{split}
\mathcal{L}_{a} &\supset \frac{1}{2}\partial^{\mu}a \partial_{\mu}a - \frac{1}{2}m_a^2 a^2 
+ \sum_q \frac{C_{qq}}{2\Lambda}(\partial^{\mu} a) \bar{q} \gamma_{\mu} \gamma_5 q + \sum_\ell \frac{C_{\ell\ell}}{2\Lambda}(\partial^{\mu} a) \bar{\ell} \gamma_{\mu} \gamma_5 \ell \\
&\quad + g^2 \frac{C_{WW}}{\Lambda} a W^{\mu \nu} \tilde{W}_{\mu \nu} + g^{\prime 2} \frac{C_{BB}}{\Lambda} a B^{\mu \nu} \tilde{B}_{\mu \nu} + g_s^2 \frac{C_{gg}}{\Lambda} a G^{\mu \nu} \tilde{G}_{\mu \nu} \; ,
\end{split}
\end{align}
where $\Lambda$ denotes the unknown scale of new physics (which may or may not be connected to the symmetry breaking scale~\cite{Alonso-Alvarez:2021ett}) and the parameters $C$ denote the Wilson coefficients of the various effective operators. For the remainder of this discussion we will set $C_{qq} = 0$ at tree level. For a more detailed discussion of the effects of ALP-quark couplings, we refer to Ref.~\cite{Bauer:2021mvw}.

At energy scales below EW symmetry breaking the most relevant interactions of ALPs with electroweak gauge bosons will then have the following form:
\begin{equation}
\mathcal{L}_{a, \text{EW}} \supset e^2 \frac{C_{\gamma\gamma}}{\Lambda} a F^{\mu \nu} \tilde{F}_{\mu \nu} + 2 e^2 \frac{C_{\gamma Z}}{\Lambda} a F^{\mu \nu} \tilde{Z}_{\mu \nu} + 4 g^2 \frac{C_{WW}}{\Lambda} a \varepsilon ^{\mu \nu \alpha \beta} \partial_{\mu} W_{\nu}^+\partial_{\alpha} W_{\beta}^- \; ,
\end{equation}
where $e = g \sin\theta_W = g^{\prime} \cos\theta_W$ and the Wilson coefficients are given by
\begin{align}
C_{\gamma\gamma} & = C_{WW} + C_{BB} \\ C_{\gamma Z} & = \dfrac{C_{WW} \cos^2\theta_W  - C_{BB} \sin^2\theta_W}{\sin \theta_W \cos \theta_W}.
\end{align}

At the one-loop level, $C_{\gamma\gamma}$ receives additional contributions from lepton and gauge boson loops, leading to
\begin{equation}
    C_{\gamma\gamma} = C_{WW} + C_{BB} + \sum_{\ell} \frac{C_{\ell\ell}}{16\pi^2} B_1(4m_{\ell}/m_a) + \frac{2 \alpha}{\pi}\frac{C_{WW}}{\sin^2 \theta_W} B_2(4m_W/m_a) \; ,
    \label{eq:Cgg}
\end{equation}
where the functions $B_{1,2}(\tau)$ are defined in eq.~(14) of Ref.~\cite{Bauer:2017ris}. The analogous one-loop contributions to the effective leptonic coupling $C_{\ell\ell,\text{eff}}$ are given in eq.~(26) of Ref.~\cite{Bauer:2017ris}, which is also implemented in our code.

\subsection{FCNC transitions}

Flavour-changing interactions are generated from the $a$-$W$ interaction at the one-loop level \cite{Izaguirre:2016dfi} and from the $a$-$g$ interaction at the two-loop level \cite{Chakraborty:2021wda}. The corresponding effective interactions take the following form:
\begin{align}
\label{eq:LFCNC}
\begin{split}
\mathcal{L}_{b \to sa} & \supset -\frac{C_{bs}}{\Lambda} (\partial^{\mu} a) \bar{s}_{\textsc{l}} \gamma_{\mu} b_{\textsc{l}} + \text{h.c.} \; , \\
\mathcal{L}_{c \to ua} & \supset -\frac{C_{cu}}{\Lambda} (\partial^{\mu} a) \bar{u}_{\textsc{l}} \gamma_{\mu} c_{\textsc{l}} + \text{h.c.} \; .
\end{split}
\end{align}
The effective FCNC couplings can be written as $C_{qq'} = C_{WW} g_{qq',W} + C_{gg} g_{qq',g}$.
The one-loop contribution from $W$ bosons is given by
\begin{align}
\label{eq:gW}
\begin{split}
g_{bs,W} &= 3 \alpha_W^2 \sum_{q = u,c,t} V_{qb} V_{qs}^{\star} F(\xi_q) \; , \\
g_{cu,W} &= 3 \alpha_W^2 \sum_{q = d,s,b} V_{uq} V_{cq}^{\star} F(\xi_q) \; , \\
\end{split}
\end{align}
where $\alpha_W = g^2/(4\pi)$, $\xi_q = m_q^2 / m_W^2$ and
\begin{equation}
F(x) = \frac{x\left[1+x(\ln x -1)\right]}{(1-x)^2} \; .
\end{equation}

The two-loop contribution $g_{bs,g}$ has been calculated for the first time in Ref.~\cite{Chakraborty:2021wda}:
\begin{equation}
    g_{bs,g} =  \frac{\alpha_s^2\alpha_W}{4 \pi}\left[f(\mu) + 2 A C_F g(\mu) + 2 B C_F \sum_{q=u,c,t} V_{qb} V_{qs}^{\star} \xi_q \right] \; 
\end{equation}
with functions $f(\mu)$ and $g(\mu)$ given in eqs.~(B6) and (B7) of Ref.~\cite{Chakraborty:2021wda}. The analogous expression for $g_{cu,g}$ is obtained by the obvious replacements.
In our code, we associate $\mu$ with the new-physics scale $\Lambda$, which can be specified by the user as well as the model-dependent parameters $A$ and $B$. For the results presented in section~\ref{sec:results}, we set $A = B = 0$, which was shown in Ref.~\cite{Chakraborty:2021wda} to yield neither overly optimistic nor overly conservative exclusions.

We emphasize that both the contribution to the effective FCNC couplings from $W$ bosons and the contribution from gluons are approximately proportional to $\xi_Q$, where $Q$ is the heaviest quark in the loop. The resulting decay rates are then proportional to $\xi_Q^2$. This leads to a strong suppression of the $c \to u+a$ transition (where $Q = b$) relative to the $b \to s+a$ transition (where $Q = t$), such that the process $B \to K^{(\star)} + a$ turns out to be experimentally more important than $D \to \pi + a$ in spite of the greater $D$ meson production cross section.

\subsection{ALP--meson mixing}

At energies below the QCD confinement scale, the ALP-gluon coupling gives rise to both kinetic and mass mixing of ALPs with the SM neutral pseudoscalars $P \in \{\pi^0,\eta,\eta'\}$. Removing this mixing requires a redefinition of the various fields:
\begin{equation}
P \to P + \theta_{aP} a
\end{equation}
with the mixing angle
\begin{equation}
\theta_{aP} = \frac{f_\pi}{f_a} \frac{K_{aP} m_a^2+ m_{aP}^2}{m_a^2-m_P^2} \; ,
\end{equation}
where $f_{\pi} = 93 \ \text{MeV}$ and $f_a = \Lambda / 32 \pi^2 \left\vert C_{gg}\right\vert$ denote the pion and ALP decay constants.\footnote{We note that our definition of $f_a$ and the ALP-meson mixing is not invariant under a chiral rotation of the SM quark fields, even though such a rotation should not affect physical observables. Rather than providing a more general treatment, we follow the common approach in the literature and assume that the chiral rotation is fixed in such a way that flavour-diagonal couplings to light quarks are absent.}

The kinetic mixing terms are given by
\begin{align}
\begin{split}
  K_{a \pi^0} &= \begin{aligned}[t]
      &-\frac{\kappa_u-\kappa_d}{2}
       \end{aligned}\\
  K_{a \eta} &= \begin{aligned}[t]
      &-\frac{(\kappa_u+\kappa_d)(\frac{1}{\sqrt{3}}\cos\theta_{\eta\eta'}-\sqrt{\frac{2}{3}}\sin\theta_{\eta\eta'})-\kappa_s(\frac{2}{\sqrt{3}}\cos\theta_{\eta\eta'}+\sqrt{\frac{2}{3}}\sin\theta_{\eta\eta'})}{2}
       \end{aligned}\\
  K_{a \eta'} &= \begin{aligned}[t]
      &-\frac{(\kappa_u+\kappa_d)(\frac{1}{\sqrt{3}}\sin\theta_{\eta\eta'}+\sqrt{\frac{2}{3}}\cos\theta_{\eta\eta'})-\kappa_s(\frac{2}{\sqrt{3}}\sin\theta_{\eta\eta'}-\sqrt{\frac{2}{3}}\cos\theta_{\eta\eta'})}{2}
       \end{aligned}
\end{split}
\end{align}
where $\theta_{\eta\eta'} \approx - 13^{\circ}$ is the $\eta$-$\eta^\prime$ mixing angle and we make the convenient choice $\kappa_q = m_q^{-1} / \mathrm{Tr} [\boldsymbol{m}^{-1} ] $ with $\boldsymbol{m} = \text{diag}\{m_u,m_d,m_s\}$ to achieve $m_{a\pi^0} = m_{a\eta_8} = 0$~\cite{Georgi:1986df}. The remaining mass mixing terms are
\begin{align}
\begin{split}
  m^2_{a \eta} &= \begin{aligned}[t]
      &-\frac{\sqrt{6}B_0 \sin\theta_{\eta\eta'}}{\text{Tr}[ \boldsymbol{m}^{-1} ]}
       \end{aligned}\\
  m^2_{a \eta'} &= \begin{aligned}[t]
      &\frac{\sqrt{6}B_0 \cos\theta_{\eta\eta'}}{\text{Tr} [ \boldsymbol{m}^{-1} ]}
       \end{aligned}
\end{split}
\end{align}
where $B_0 = m_{\pi^0}^2 / (m_u+m_d)$.

The ALP-gluon coupling as well as ALP-meson mixing give rise to further contributions to the effective ALP-photon coupling, which is found to be~\cite{GrillidiCortona:2015jxo}
\begin{equation}
C_{\gamma \gamma,\text{eff}} = C_{\gamma\gamma} - 1.92 C_{gg} + \tilde{\alpha}_s(m_a^2)\mathcal{F}_{\text{VMD}} \sum_{P}\frac{\Lambda}{16 \pi^2 f_{P}} \theta_{aP}
 \label{eq:Cggeff}
\end{equation}
with $f_{\eta} \approx 93 \ \text{MeV}$ and $f_{\eta'} \approx 73 \ \text{MeV}$.
The functions $\mathcal{F}_{\text{VMD}}$ and $\tilde{\alpha}_s$, defined in eqs.~(\ref{eq:Fvmd}) and (\ref{eq:alphas}) respectively, allow us to extrapolate the mixing effects for $m_a > m_{\eta'}$.

\section{Calculation of decay widths}
\label{sec:decay_calculation}

Based on the effective couplings derived in appendix~\ref{sec:calculations} we can calculate the various decay widths relevant for this study. 

We begin by noting that in all cases of interest the decaying particle has spin $0$ and hence the squared matrix element does not depend on the phase space parameters, leading to an isotropic decay. Hence no information on the decay kinematics is lost when performing the phase space integration and it is sufficient to quote the corresponding decay width.

The width of the decay $a \to \gamma \gamma$ is given by
\begin{equation}
\Gamma_{a \to \gamma\gamma} = \frac{C^2_{\gamma\gamma,\text{eff}}}{\Lambda^2} \frac{e^4 m_a^3}{4 \pi}
\end{equation}
with an effective ALP-photon coupling $C_{\gamma\gamma,\text{eff}}$. 
For the ALP leptonic decays $a \to \ell\ell$ ($\ell = e,\mu$) with decay width
\begin{equation}
\Gamma_{a \to \ell\ell} = \frac{C^2_{\ell\ell,\text{eff}}}{\Lambda^2} \frac{m_\ell^2}{8 \pi} \sqrt{m_a^2 - 4 m_\ell^2} \, \;
\end{equation}
where $m_\ell$ denotes the lepton mass and $C_{\ell\ell,\text{eff}}$ 
the effective leptonic coupling.

Other possible two-body decays for ALP masses up to $3\,\mathrm{GeV}$ are the hadronic decays $a \to \rho\rho$, $a \to \omega\omega$, $a \to K^{\star} \bar{K}^{\star}$ and $a \to \phi\phi$, which (in the absence of direct couplings to quarks) have a decay width proportional to $C_{gg}^2$. Since these final states are not \textit{detector-stable} for fixed-target experiments, we do not calculate them explicitly but we account for them in the decay widths for three-body decays. The corresponding decay widths used in this work employ amplitudes derived in Ref.~\cite{Aloni:2018vki} using chiral perturbation theory and vector meson dominance.

Furthermore, we are also interested in the production of ALPs via flavour-changing two-body decays. The decay widths for the rare decays $B \to K^{(\ast)} + a$ can be calculated in terms of the \textit{flavour-changing} effective couplings $C_{bs}$:
\begin{align}
\label{eq:Bwidth}
\begin{split}
\Gamma_{B \to K + a} &=
\begin{aligned}[t] &\frac{|C_{bs}|^2}{\Lambda^2}\frac{1}{64 \pi m_B^3} (m_B^2-m_K^2)^2 f_0^2(m_a) \lambda^{1/2}(m_B,m_K,m_a) \end{aligned}\\
\Gamma_{B \to K^{\star} + a} &=
\begin{aligned}[t] &\frac{|C_{bs}|^2}{\Lambda^2}\frac{1}{64 \pi m_B^3} A_0^2(m_a) \lambda^{3/2}(m_B,m_{K^{\star}},m_a) \; ,\end{aligned}
\end{split}
\end{align}
where $\lambda(x,y,z) = \left[x^2-(x+z)^2\right]\left[x^2-(y-z)^2\right]$. Since we apply this result on a broad range of ALP masses, we use the following parametrization of the form factors~\cite{Ball:2004ye,Ball:2004rg}:
\begin{align}
f_0(q) & = \frac{0.330}{1-q^2/37.46} \; ,\\
A_0 (q) & = \frac{1.364}{1-q^2/m_B^2} - \dfrac{0.990}{1-q^2/36.78 \ \text{GeV}^2} \; ,
\end{align}
which depend on the momentum transfer $q$.

In complete analogy, we find for the decay $D \to \pi + a$: 
\begin{equation}
\label{eq:Dwidth}
\Gamma_{D \to \pi + a} =
\frac{|C_{cu}|^2}{\Lambda^2}\frac{1}{64 \pi m_D^3} (m_D^2-m_\pi^2)^2 f_0^2(m_a) \lambda^{1/2}(m_D,m_\pi,m_a) \; , \end{equation}
where we use $f_0(0) = 0.612$~\cite{Lubicz:2017syv} (see also Ref.~\cite{MartinCamalich:2020dfe}). To account for the $q$-dependence we use an analogous parametrization as above:
\begin{equation}
f_0 (q) = \frac{f_0(0)}{1-q^2/m^2_{\text{fit}}}
\end{equation}
For fitted $m^2_{\text{fit}} = 6.46\,\mathrm{GeV}^2$ we reproduce the mean values of Ref.~\cite{Lubicz:2017syv} with a relative error $< 1\%$, which is sufficient for our purpose. 

\section{Brief description of the \texttt{ALPINIST} code}
\label{sec:ALPINIST}

\begin{figure}[t]
\begin{center}

\includegraphics[scale=1]{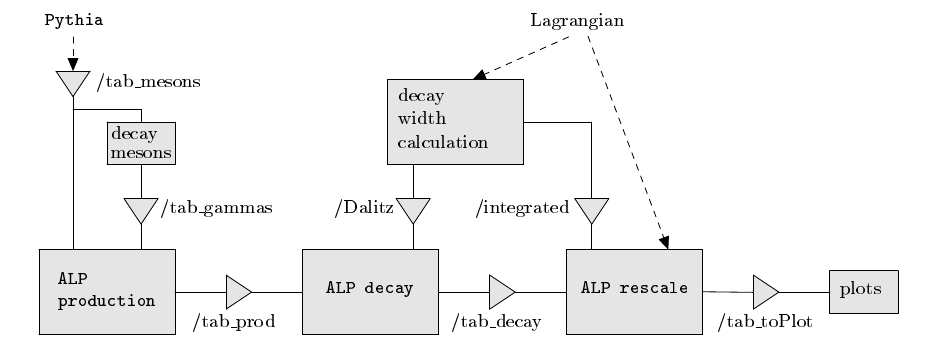}
\caption{\label{fig:code_structure} Diagram representing the internal structure of the \texttt{ALPINIST} framework.}
\end{center}
\end{figure}

Since the MC simulations of ALP production and decay are computing power demanding, our goal is to avoid as much as possible re-doing these simulations for each change in the model-dependent parameters. Thus we separate the code into several standalone modules, each of which processes the input tabulated data and passes the output data, again in the form of a table, to the next module in the sequence. This sequence is shown in the form of a diagram in figure~\ref{fig:code_structure}, where the code-modules are represented by square boxes and the tabulated data by triangular boxes pointing in the direction of the data flow.

\begin{itemize}
\item The \texttt{ALP production} module uses \texttt{Wolfram Mathematica} \cite{Mathematica}. It corresponds to section \ref{sec:production}, i.e.\ it generates the ALP distribution table for each production mode and for each experiment for a fixed value of the model-dependent parameters. For the ALP production from mesons it reads directly the output from the \texttt{PYTHIA} simulations. For the production from on-shell photons, it reads the photon distributions generated by a separate module, which simulates $2\gamma$ decays of $\pi^0$, $\eta$ and $\eta^{\prime}$ mesons. This module uses the \texttt{ROOT} framework \cite{ROOT}.
\item The \texttt{ALP decay} module also uses the \texttt{ROOT} framework libraries. It contains the main MC simulation described in sections \ref{sec:decay} and \ref{sec:exp}, i.e.\ it handles the ALP decay kinematics for various decay modes in the conditions of the specific experiment. For the three-body decays it also requires the Dalitz plot densities, which are generated according to the specific Lagrangian (eq. (\ref{eq:Lgen}) in this case) by a separate module. 
\item The \texttt{ALP rescale} module requires only \texttt{Python} \cite{10.5555/1593511} and the input tables to be run. For the set of model-dependent parameters given by the user it generates the table with the number of signal events in the common $m_a$-$C_a$ plane used for plotting the exclusion limits. The equations used for the calculations can be found in appendices \ref{sec:calculations}, \ref{sec:decay_calculation} as well as the references therein.
\end{itemize}

The \texttt{ALP rescale} module is the most relevant one for most users and thus is provided publicly at \url{https://github.com/jjerhot/ALPINIST}  together with its input tables. Other modules may be needed for potential updates of experimental conditions and for including new experiments, production or decay modes to the simulation. Access to the full repository can be obtained upon request from the authors. The specific instructions for using of each module can be found in the respective \texttt{README.md} files.

\section{Treatment of kinematics for the mixing production}
\label{ap:mixing_kinematics}

The mixing production serves as an approximation to the various processes in which an ALP can be produced. The total yield of produced ALPs can be derived from the yield of the neutral pseudoscalar mesons and the respective mixing angles with the ALP. This approach, however, does not say anything about the change of the kinematics when the ALP is produced instead of the meson. The general minimal adjustment, which can be applied irrespective of the specific process which lead to the ALP production, is the change of the kinematics when the mass of the original pseudoscalar is changed. Since the center-of-mass frame of the processes that lead to the meson production is unknown, we choose a common frame for the mass adjustment to be the $p$-$p$ collision \textit{cm frame}. We emphasize that there is a certain level of arbitrariness in this choice, with the obvious alternative being the laboratory frame. However, we expect that the $p$-$p$ \textit{cm frame} leads to a more accurate description for highly-energetic ALPs, where only a smaller portion of the available energy is spent on the ALP production and the majority of the energy is carried in the ALP boost. These highly boosted ALPs are particularly important for the sensitivity of beam dump experiment searches.

\begin{figure}[t]
\begin{center}
\includegraphics[scale=.67]{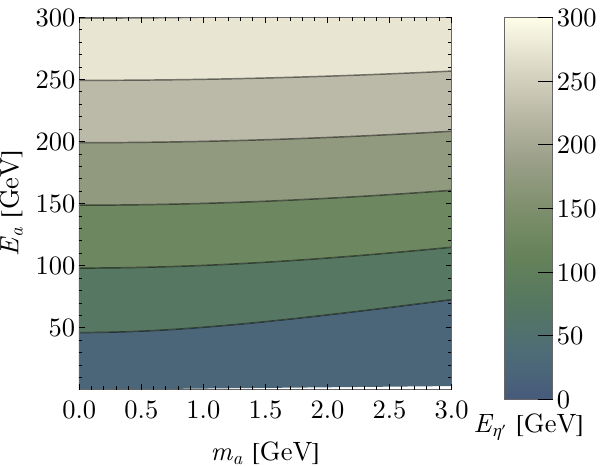}
\includegraphics[scale=.67]{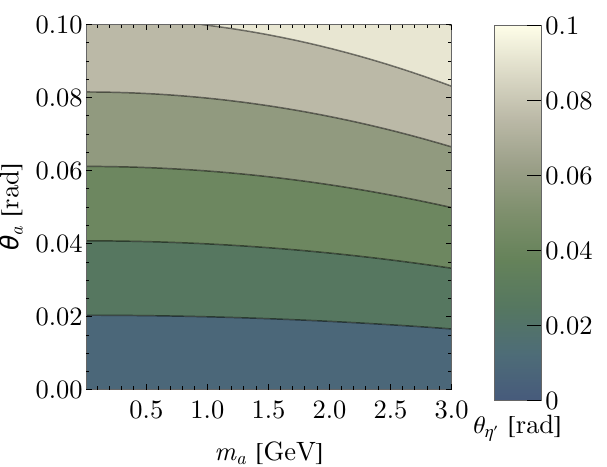}
\includegraphics[scale=.67]{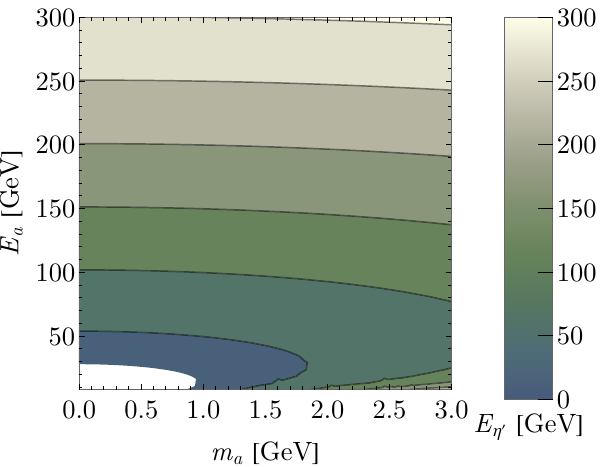}
\includegraphics[scale=.67]{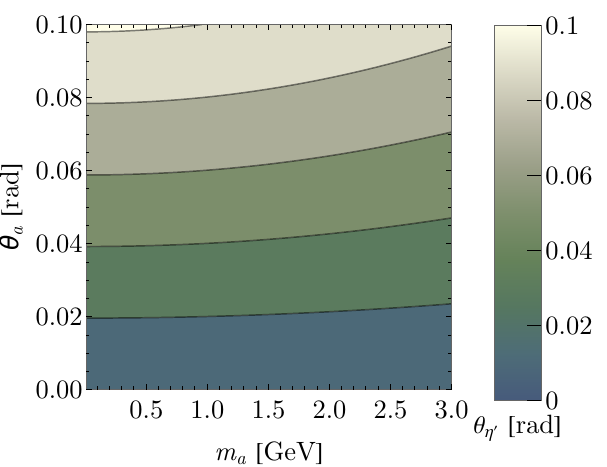}
\caption{\label{fig:mesonPfixed} Dependence of the energy (left) and the angle of the ALP (right) on the ALP mass in the laboratory frame for the case of the fixed momentum (upper panel) or fixed energy (lower panel) in the $p$-$p$ \textit{cm} frame when the original particle was the $\eta^{\prime}$ meson, produced by a $400 \; \mathrm{GeV}$ proton beam. The angular distribution is evaluated for the case $E_a = 100 \; \mathrm{GeV}$.}
\end{center}
\end{figure}

The \textit{cm} momentum of a particle with mass $m$ is then simply found as
\begin{equation}
p_{\text{cm}}(p_{\text{lab}},m) = \gamma_{\text{beam}} p_{\text{lab}} - \gamma_{\text{beam}} \beta_{\text{beam}} \sqrt{p_{\text{lab}}^2 + m^2}
\end{equation}
and analogous relations can be used for the $E_{\text{cm}}$ and, by inversion, for $p_{\text{lab}}$ and $E_{\text{lab}}$. By adjusting the particle mass one necessarily violates the momentum or energy conservation (or both) in the original process. In our simulation we fix the momentum of the ALP in the  $p$-$p$ \textit{cm frame} to be the same as the momentum of the original meson. The energy and angle of the ALP $a$ in the laboratory frame then translates to the energy and angle of the original meson $P$ in the laboratory frame as
\begin{align}
\label{eq:mesonPfixed}
\begin{split}
E_P \big|_{p_\mathrm{fixed}} &= \sqrt{p_{\mathrm{lab}}\left(p_{\mathrm{cm}}\left(p_a,m_a\right),m_P\right) + m_P^2} \, ,\\
\theta_P \big|_{p_\mathrm{fixed}} &= \arcsin \left[ \sin(\theta_a) \frac{p_a}{p_{\mathrm{lab}}\left(p_{\mathrm{cm}}\left(p_a,m_a\right),m_P\right)} \right] \; .
\end{split}
\end{align}
The plotted distributions for the case of $P = \eta^{\prime}$ can be found in the upper row of figure~\ref{fig:mesonPfixed}. 

If one chooses to fix the ALP energy to the meson energy in the $p$-$p$ \textit{cm frame}, one obtains in complete analogy the following relations:
\begin{align}
\label{eq:thetaMeson}
\begin{split}
E_P \big|_{E_\mathrm{fixed}} &= E_{\mathrm{lab}}\left(E_{\mathrm{cm}}\left(p_a,m_a\right),m_P\right) \, \\
\theta_P \big|_{E_\mathrm{fixed}} &= \arcsin \left[ \sin(\theta_a) \frac{p_a}{\sqrt{E_{\mathrm{lab}}\left(E_{\mathrm{cm}}\left(E_a,m_a\right),m_P\right)^2 - m_P^2}} \right] \; .
\end{split}
\end{align}
The resulting distributions are  plotted in the bottom row of figure~\ref{fig:mesonPfixed}. By comparing the two distributions, we can see that even in the borderline case of choosing the $p$-$p$ collision as a universal \textit{cm} frame, we can get a significantly different $E_a$-$\theta_a$ distribution solely by choosing a different method of fixing the kinematics in the \textit{cm frame}. This choice, however, impacts mostly ALPs with a small boost (large masses and low energies), which are of limited relevance for beam dump experiments.

\end{appendix}

\input{ALPINIST.bbl}

\end{document}

%% file: ALPINIST.bbl
\providecommand{\href}[2]{#2}\begingroup\raggedright\endgroup